\documentclass[manuscript]{aastex}
\begin{document}
\newcommand{\etal}{et~al.}
\tighten
\title{ A Spectroscopic Survey of Subarcsecond Binaries in the Taurus-Auriga Dark Cloud
with the Hubble Space Telescope \altaffilmark{1}}

\author{
	Patrick Hartigan \altaffilmark{2}, 
	and Scott J. Kenyon\altaffilmark{3,4}
	}

\vspace{1.0cm}

\altaffiltext{1}{Based on observations made with the NASA/ESA
{\it Hubble Space Telescope}, obtained at the Space Telescope Science
Institute, which is operated by the Association of Universities for
Research in Astronomy, Inc., under NASA contract NAS5-26555.}

\altaffiltext{2}{Dept. of Physics and Astronomy, Rice University, 
6100 S. Main, Houston, TX 77005-1892, USA}

\altaffiltext{3}{Center for Astrophysics, 60 Garden St.,
Cambridge MA 02138, USA} 

\altaffiltext{4}{Visiting astronomer, Kitt Peak National Observatory}

\begin{abstract}

We report the results of a spectroscopic survey of 20 close T~Tauri binaries
in the Taurus-Auriga dark cloud where the separations between primaries and their
secondaries are less than the typical size of a circumstellar disk around a young star.
Analysis of low-resolution and medium-resolution STIS spectra yields
the stellar luminosities, reddenings, ages, masses, mass accretion rates, IR excesses, and
emission line luminosities for each star in each pair.
We examine the ability of IR color excesses, H$\alpha$ equivalent widths,
[O~I] emission, and veiling to distinguish between weak emission and classical
T Tauri stars.  Four pairs have one cTTs and
one wTTs; the cTTs is the primary in three of these systems. 
This frequency of mixed pairs among the close T~Tauri binaries is similar to the
frequency of mixed pairs in wider young binaries.
Extinctions within pairs are usually similar; however,
the secondary is more heavily reddened than the primary in some systems,
where it may be viewed through the primary's disk. 
Mass accretion rates of primaries and secondaries are strongly correlated, and
H$\alpha$ luminosities, IR excesses, and ages also correlate within pairs. 
Primaries tend to have somewhat larger accretion rates than
their secondaries do, and are typically slightly older than their secondaries according
to three different sets of modern pre-main-sequence evolutionary
tracks.  Age differences for XZ~Tau and FS~Tau, systems embedded in
reflection nebulae, are striking; the secondary in each pair is less massive
but more luminous than the primary.  The stellar masses of the UY~Aur and
GG~Tau binaries measured from their rotating molecular disks are about
30\%\ larger than the masses inferred from the spectra and evolutionary
tracks.  This discrepancy can be resolved in several ways, among them 
a 10\%\ closer distance for the Taurus-Auriga dark cloud.


\end{abstract}

\section{ INTRODUCTION  }

A large body of work in the last two decades has established that binaries
with active accretion disks are the most common way that stars form.
Surveys of regions of star formation using speckle images \citep{ghez93,lein93}
and lunar occultations \citep{simon99} have established that binaries are
ubiquitous among the youngest stars. Direct imaging \citep{mcc96,padgett99}
and surveys of infrared excesses
\citep{haisch01} have discovered circumstellar disks around many
newly-formed stars.  Most young circumstellar disks accrete
onto their central stars, as evidenced by inverse P-Cygni line profiles
\citep{edw94} and excess emission from the accretion hot spot at optical
and ultraviolet wavelengths \citep[e.g.][]{hartigan95,gullbring98}.

To understand how stars form we must address
several fundamental questions concerning binaries and their disks.
One of the most basic of these is whether or not the components
of a young binary have the same age. If capture plays a role in binary formation
then a range of ages should occur within pairs, while if the secondary forms
out of a disk around the primary then the secondary should be younger than the primary.
If the relative location of primaries and their secondaries is consistent throughout
the HR-diagram, it may be possible to measure the location of isochrones that must
be reproduced by theoretical models of young stars.

While most pairs generally fall along isochrones derived from theoretical
pre-main-sequence evolutionary tracks \citep{hartigan94,brandner97,white99},
uncertainties in the effective temperature scale at the lowest masses and
difficulties in estimating stellar luminosities, especially when individual
spectra of the two stars are lacking, have made it difficult to compare ages
within pairs. In some pairs infrared companions exist that are much
more obscured, and therefore possibly significantly younger, than
their optically visible primaries. However, extinction does not
necessarily correlate inversely with age -- highly embedded companions may
be obscured only temporarily by an episode of infalling envelope material. Hence,
these infrared companions are not necessarily younger simply because they
are more reddened \citep{koresko97}.  Any secondary that is more luminous
than its primary must also be younger, because isochrones 
always slope down and to the right in an HR diagram for pre-main-sequence stars.

Other important issues are whether the masses, spatial extents, lifetimes,
and accretion histories of primary circumstellar disks differ from those that
surround the secondaries.  Recent studies of the frequency of disks in young binaries
have concluded that mixed pairs -- those where one star is a classical T~Tauri star (cTTs;
stars with circumstellar disks) and the other a weak-lined T~Tauri star (wTTs; stars without detectable
disks) -- are rare \citep{prato97,duchene99}. Stars with the highest accretion rates tend
to be more massive (i.e., primaries, on average, accrete more rapidly than secondaries do),
but there is no obvious correlation between mass accretion rates
and separations \citep[][hereafter WG01]{wg01}.

How disks interact in binary systems when the distance between the stars
is smaller than the typical size of a disk, $\sim$ 100~AU, is a topic of much current research.
Models of these close binaries typically surround the stars with a circumbinary disk,
as has been imaged with adaptive optics and with HST around some systems \citep{close98,silber00}. 
In DQ~Tau, it appears that the eccentric orbit of the binary leads to cyclic pulses of accretion
from the circumbinary disk onto the stars \citep{mathieu97,basri97}. In such systems the less massive star may
accrete faster than the more massive star does \citep{arty96}, a prediction that can in principle
be tested observationally.

To make progress in these areas one must measure the spectral types, accretion rates,
luminosities, emission line fluxes, and reddenings of each primary and secondary independently, which
requires individual spectra and photometry for each star. The stellar luminosities and 
effective temperatures, when compared with predictions of published
pre-main-sequence evolutionary tracks, provide the masses and ages of each star.
If the primaries are, on average, coeval with their secondaries,
it should be possible to test the accuracy of the theoretical tracks, 
which have been undergoing significant revisions as the assumption of grey photospheres
has been relaxed \citep{siess00,baraffe98,palla99}.
It is also possible to test the evolutionary
tracks if there is an independent measurement of the stellar masses from orbital motion
\citep{ghez95,stef01} or from Keplerian rotation of the molecular disk \citep{guill99,simon00}.

In this study we focus on 20 close pairs where the projected separations are $\lesssim$ 100~AU.
Even in the Taurus-Auriga complex, the closest significant population of cTTs, these size scales
are all subarcsecond and most cannot be resolved from the ground. Our spectroscopic study extends
the work of WG01, who mainly used photometric HST observations to investigate close binaries.
By combining both low-resolution and medium-resolution spectra together we can measure spectral
types, reddenings, accretion and emission line properties of each component of each pair
accurately. Together with the WG01 photometry, the new spectra give the best overview to date of
how young close binary systems interact with their disks. 

\section{ OBSERVATIONS AND DATA REDUCTION}

Spectra of 20 close T~Tauri binaries were observed using the CCD detector attached to
the STIS spectrograph on the Hubble Space Telescope between 26 Nov 1998 and 22 Jan 2001. 
The medium-resolution spectra taken with the G750M grating with central wavelength
6252\AA\ have a resolution of R $\sim$ 7000, a dispersion of 0.56 \AA\ per pixel,
and range from 5950\AA\ -- 6550\AA . Similarly, the low-resolution spectra that used
the G750L grating at central wavelength 7750\AA\ have a resolution of R $\sim$ 900, a
dispersion of 4.9\AA\ per pixel, and cover 5270\AA\ -- 10260\AA . The central wavelengths
are optimal for measuring TiO bands in reddened young stars.

Ideally one would align the 52$^{\prime\prime}$ slit of STIS
with the stars in each pair, but for many targets this was not possible
owing to severe scheduling constraints imposed by the proximity of the targets to
the ecliptic. In most cases a 0.2$^{\prime\prime}$
slit was used, but five pairs (DD~Tau, FQ~Tau, FV~Tau, UZ~Tau~W, and V999~Tau)
required the 0.5$^{\prime\prime}$ slit to allow the observations to be done
over a larger range of position angles, and for V927~Tau (Lk H$\alpha$ 331), the 
$36^{\prime\prime}\times 0.6^{\prime\prime}$ slit was the only one that
allowed the observations to be done. The spectral resolution is the 
same to within measurement error through these three slits.

Reduction of the spectra followed the standard pipeline procedure with
several modifications. Sporadic dark counts on the STIS CCD were the dominant
source of noise in most spectra. Subtracting a `day dark' improved the situation
somewhat, but many hot pixels remained. To clean the data we developed a procedure
where statistics measured over a box are compared with the pixel value at the
center of the box to determine if the value is erroneously high or low. Real
absorption or emission lines will display the PSF of the system in the spatial
dimension, while hot pixels will not. Through trial and error and 
individual inspection of each two-dimensional image one can effectively eliminate
all the hot pixels in this manner without affecting any real features. The low-resolution
data had to be corrected for fringing using flats taken for this purpose.
The correction procedure reduced the amplitude of the fringing across the entire spectrum
to less than 5\%, a significant improvement over the 30\%\ -- 40\%\ amplitudes
present in the reddest portion of the spectra before the fringe corrections. 
Fringing varies spatially over only a few pixels, so it is necessary to 
correct the secondary separately from the primary. Fringing corrections were negligible
for wavelengths $\lesssim$ 8000\AA .

Flux levels of the medium-resolution and low-resolution spectra of the same star
taken within the same orbit agree on average within 8\% , so the absolute flux calibrations
and extractions agree to this level.  In all cases the slopes of the medium-resolution and
low-resolution spectra are essentially identical.  Errors in relative flux
calibrations at different wavelengths across a single spectrum are dominated by PSF
variations in STIS, especially at the longest wavelengths in the low-resolution
spectra, and should be $\lesssim$ 5\% . 
The effect of the PSF on our medium-resolution spectra is $<$ 1\% . The PSFs were corrected
according to the tables appropriate for the slit used in the observations.

When the pairs are separated by more than about 0.3$^{\prime\prime}$, it is
possible to extract each spectrum from the 2D data by simply summing the
appropriate rows and applying the appropriate flux correction with wavelength
given the extraction aperture (the PSF of STIS develops an extended tail in the
near-IR that is corrected in the flux calibration, which depends on the aperture size). 
However, when the separation is less than 0.3$^{\prime\prime}$ the PSFs of the
primary and secondary overlap, and must be deblended. The deblending was done by
first extracting the spectrum of primary+secondary with a large aperture, and
then determining the ratio of the primary to secondary flux at each wavelength
by fitting a gaussian to the location of the primary and secondary, which are known
empirically from the acquisition image taken immediately before the spectrum.
The result proves to be essentially independent of the choice of the fitting function because
the goal is only to determine the ratio of primary to secondary fluxes. 
By combining the ratio spectrum with the combined spectrum we obtained individual spectra
of the primaries and secondaries.

We define the `primary' as the brighter star at 7000\AA . In the case of FQ~Tau, where
the fluxes of the two stars are nearly equal at 7000\AA , we
take the bluer star, which has an earlier spectral type, to be the primary.
With these definitions the secondary can be more luminous overall, especially if it is very red,
but the primary always turns out to be the most massive star in the pair for the
objects in our sample.

Position angles and separations of the pairs reported in Table~1
were measured from the unfiltered 100x100 CCD acquisition images (Fig.~1)
obtained with STIS. Uncertainties in the PA and separation are dominated
by centroiding errors, which are typically 0.1 -- 0.2 pixels.
By comparison, uncertainties in the plate scale of STIS
(0.05072 arcseconds per pixel) and orientation of HST are negligible.
All of the pairs with separations $\lesssim$ 0.3$^{\prime\prime}$
(DF~Tau, FO~Tau, GG~Tau, FS~Tau, and IS~Tau) have position angles that
differ by 10 degrees or more from those in the literature, while 
the discrepancies for more widely separated pairs are all smaller.
The PA of a circular orbit at the distance to Taurus with a separation of
0.2 arcsecond and a combined mass of 1 M$_\odot$ 
varies by 10 degrees in about 4 years.  The PA of the closest binary,
DF Tau (separation 0.099$^{\prime\prime}$, changed by $\sim$ 30 degrees from the value reported by
\cite{ghez97} from previous HST images taken in July of 1994.

\section{SPECTRAL TYPES, REDDENING, LUMINOSITIES, AND MASS ACCRETION RATES}

\subsection{Spectral Diagnostics}

Deriving the physical properties of the inner disk and stellar
photosphere of the T Tauri star requires estimates of the reddening, 
spectral type, and veiling.  Because veiling fills in
absorption features and makes the broadband colors of the
underlying stellar photosphere more blue, the veiling must be measured 
before assigning a stellar spectral type and deriving the 
reddening.  To measure veiling, we approximate the
underlying stellar spectrum with a `template spectrum' chosen
to match the stellar photosphere as closely as possible. 
We then adopt a spectral shape for the veiling emission (section 3.2)
and use a least-squares minimization to solve for the amount of veiling
and reddening for that template. We then repeat this procedure for several
templates to find the best solution.  For the template spectra, we 
choose weak emission T Tauri stars acquired with the 
same instrumentation.  The best templates are those that leave no
residual spectral features when the model (template plus veiling) is
subtracted from the object.

To make first estimates for the stellar spectral type and reddening,
we measure continuum magnitudes and spectral indices for strong
absorption lines.  For comparison, we derive continuum magnitudes and absorption indices 
for nearby stars with known spectral types \citep{kir91}.  \citet{har91} 
describes ground-based spectrophotometry of these stars.  Briefly, the
data were acquired with the Intensified Reticon Scanner (IRS) and white
spectrograph at the No. 2 0.9-m telescope at Kitt Peak National Observatory.
The spectra have a resolution of $\sim$ 10 \AA~and cover 5700--8400 \AA.
The photometric calibration of these data has an uncertainty of $\pm$ 0.03 
mag based on observations of standard stars and repeat observations of
program stars.  

Continuum magnitudes at 5400 \AA\ and at 7035 \AA~are
normalized to the zero-point of the V magnitude scale, 
$m_{\lambda}$ = $-2.5~{\rm log}~ F_{\lambda}$ $-$21.1, 
where $F_{\lambda}$ is the average flux in a 30 \AA~passband 
centered at wavelength $\lambda$.  Absorption indices for
CaH $\lambda$6975, Ti~I $\lambda$7358, 
the Na~I $\lambda\lambda$ 5890, 5896 doublet, Ca II $\lambda$8542, 
and two TiO bands at $\lambda\lambda$ 6225, 7675 are derived from
the STIS spectra using SBANDS within IRAF.  

Fig.~2 (see also \citet{ken01}) shows that the graph of CaH vs. TiO $\lambda$7675
provides a reasonable first estimate for the 
spectral types of pre-main sequence stars because the ratio of these
indices is insensitive to gravity. The 
$m_{5400}$--$m_{7035}$ color then provides an initial estimate for 
the reddening of each star.  For wTTS, the $A_V$ derived from 
$m_{5400} - m_{7035}$ agrees with estimates derived from V--R or J--K 
to $\pm$ 0.1--0.2 mag.  For cTTS, the derived $A_V$ is sensitive to 
the veiling.  We use the veiling model to improve reddening and spectral
type estimates as we derive the veiling.

Table 2 compiles the equivalent widths for the brightest emission lines. Uncertainties
in the location of the continuum, measured by fitting the continuum at
different levels, dominate the errors in the tabulated values for the 
emission lines.

\subsection{Veiling Model}

We follow previous studies and assume that the observed spectrum of a TTS,
$F_{\lambda,obs}$, is the arithmetic sum of a photospheric spectrum 
$F_{\lambda,\star}$ and a veiling spectrum $F_{\lambda,veil}$,
\begin{equation}
F_{\lambda,obs} = a_{veil} F_{\lambda,\star} + F_{\lambda,veil} 
\end{equation}
where $a_{veil}$ is a constant that represents the ratio of the photospheric
flux of the object to that of the template.
The photospheric `template' spectrum is the observed spectrum of a wTTS,
$ F_{\lambda,\star} \equiv F_{\lambda,wTTS}$. We test this assumption
by measuring the veiling of a wTTS using another wTTS with the same 
spectral type as the template.  Stars with negligible $F_{\lambda,veil}$
are reasonable templates.

To solve for the veiling spectrum, we adopt a model for the veiling
and use a minimization procedure to match the model to the data.  
\citet{har89} assume a discrete model, where the veiling is a constant
$C_n$ in a bandpass of width $\Delta \lambda$ centered at $\lambda_n$.
They use a least-squares technique to find the best $C_n$ for a set of
$\lambda_n$ defined by the discrete orders of an echelle spectrograph; 
\citet{har91} combined the $C_n$ with spectrophotometry to derive the 
veiling spectrum \citep[see also][]{bas90}.  Here, we adopt a 
linear model for moderate resolution STIS spectra,
\begin{equation}
F_{\lambda,veil} = F_{ref,veil} + b_{veil} (\lambda - \lambda_{ref}) ~ ,
\end{equation}
where $F_{ref,veil}$ is the veiling flux at the reference wavelength
$\lambda_{ref}$ and $b_{veil}$ is a constant.  This approach avoids
choosing bandpasses from a continuous spectrum and provides a good
way to measure the variation of veiling with wavelength for low- and
medium-resolution spectra like those reported here.  In the low
resolution spectra, we adopt a quadratic model,
\begin{equation} 
F_{\lambda,veil} = F_{ref,veil} + b_{veil} (\lambda - \lambda_{ref}) ~ + 
c_{veil} (\lambda - \lambda_{ref})^2,
\end{equation}
where $c_{veil}$ is another constant.  

With these definitions, the veiling $r_{\lambda}$ \citep{har89} is
a simple ratio of the veiling model to the template spectrum,
\begin{equation}
r_{\lambda} = a_{veil}^{-1} F_{\lambda,veil} / F_{\lambda,\star} ~ .
\end{equation}
Averaging this quantity over bandpasses of width $\Delta \lambda$ 
yields $R_{\lambda}$, the veiling averaged over a bandpass \citep{har89}.

We use a downhill simplex method \citep[amoeba algorithm][]{pre86} 
to find the best veiling parameters for each TTS spectrum.  The
algorithm minimizes the function
\begin{equation}
\chi^2 = \sum_{i=1}^{N} \left ( \frac{F_{i,obs} - (F_{i,\star} + F_{i,veil})}{\sigma_i} \right )^2 ~ ,
\end{equation}
where $N$ is the number of discrete wavelengths and $\sigma_i$ is
an error estimate.  The `residual spectrum' is
\begin{equation}
F_{\lambda,res} = F_{\lambda,obs} - (F_{\lambda,\star} + F_{\lambda,veil})
\end{equation}
where $F_{\lambda,veil}$ uses the best-fit model parameters.
To avoid problems with wavelength solutions and flux calibrations at 
the ends of each spectrum, we clip 10--15 pixels off the end of each 
spectrum when building the model.  To estimate the noise, we add in 
quadrature the photon noise, the calibration noise, and the 
`shifting noise' defined below.  Given starting values for the unknown 
parameters, the amoeba algorithm finds the values for $a_{veil}$, 
$b_{veil}$, and $c_{veil}$ which minimize $\chi^2$ for an input 
template spectrum and $A_V$.  In our implementation, we make several
starting choices to verify the minimum; from several templates, we
select the one which minimizes $\chi^2$ and the slope of the residual
spectrum.

To derive error estimates for the model parameters, we use a Monte
Carlo technique based on the flux error estimate $\sigma_i$.  For each 
of $10^3$ to $10^4$ Monte Carlo trials, we add noise to each point of 
the observed spectrum and repeat the amoeba solution.  The noise
model is $F_{i,noise} = \sigma_{i} g_i$, where $g_i$ is a normally 
distributed deviate with zero mean and unit variance \citep{pre86}.
The observed flux for these trials is
\begin{equation}
F_{i,obs}^{\prime} = F_{i,obs} + \sigma_{i} g_i ~ .
\end{equation}
This procedure yields a range of solutions for $a_{veil}$, $b_{veil}$, 
and $c_{veil}$; we adopt the first and third quartiles of this range
as the 2$\sigma$ error in the derived parameters.

To minimize the residuals due to emission lines and differences in
wavelength solutions between the observed and template spectra, we
shift the wavelength scale of the template spectrum.  Before running 
the algorithm, we cross-correlate the observed and template spectrum 
to derive the wavelength shift. The wavelength shift has an accuracy 
of $\sim$ 0.1 \AA~or better for medium resolution spectra and $\sim$ 
0.3 \AA~or better for low resolution spectra.  Emission lines and
veiling often prevent better measurements of the wavelength shift.
We use the polynomial 
interpolation routine POLINT \citep{pre86} to transform the template 
spectrum to the wavelength scale of the observed spectrum. Tests with 
different interpolation algorithms verify that a 14th order polynomial provides flux 
conservation to better than 0.1\% over the entire wavelength scale.  
The shifting noise in the flux is the rms difference between 
transformed spectra using 12th order and 14th order polynomials.
After shifting the template spectrum, the amoeba algorithm derives
best model parameters and 1$\sigma$ errors.  We then repeat this
procedure for different wavelength shifts with a range of $\pm$ 1 \AA~for
medium resolution spectra and $\pm$ 10 \AA~for low resolution spectra.
Repeating the amoeba algorithm and Monte Carlo trials for different
wavelength shifts yields a robust minimum in $\chi^2$ and provides
better error estimates in the model parameters.

With low and medium resolution spectra for each pair of TTS, we
iterate on the input template and reddening to find the best
solution for both spectra.  We derive a best reddening by
requiring that the residual spectrum for the low resolution 
data has zero slope over $\lambda\lambda$6000--8500.  Template
mismatch prevents us from using the full wavelength range of the
low resolution data for some TTS, and fringing corrections introduce 
uncertainties at the longest wavelengths. In some cases, $\chi^2$ does
not provide a good discriminant between templates. In those cases,
we choose the best template by minimizing the residuals across
prominent absorption features such as TiO.
In some cases the spectral type of the object lies
between those of two standards; in these cases we choose the standard that
produces the best residuals and flattest veiling spectrum to display in
Figs.~3 and 4, and average the veiling and reddening results for the object
obtained using the two templates for use in the Tables and other Figures. 

Tests with different veiling models demonstrate the validity of
our veiling model for these data.  Constant veiling models for
medium resolution data and linear models for low resolution data
produce poor results and marginal constraints on the underlying
spectral type and the veiling.  Higher order polynomials do not 
improve the $\chi^2$ per degree of freedom of the fits significantly.
Trials with more physical veiling models do not improve the fits or
are unsatisfactory.  For low resolution data, blackbody models are 
similar to the polynomial models and often do not provide a useful
blackbody temperature.  The medium resolution data provide no 
constraint on the blackbody temperature.  Without measurements
covering the Balmer jump at 3646 \AA, optically thin models or slab 
models \citep{har91,val93,gullbring98} provide fewer constraints than blackbodies.
Because these more physical models often fail completely to
constrain the veiling, we do not report results for these trials.

Figure 3 shows results for the medium resolution spectra.  Each panel
of the figure includes the dereddened spectrum (top curve), the
spectrum of the normalized template star (middle curve), and the
residual spectrum (lower curve).  When the residual spectrum is 
zero at all wavelengths, the veiling spectrum is the difference
between the observed spectrum and the normalized template spectrum.
In stars with little or no veiling, spectra of the veiling emission
and the photosphere nearly coincide.
Aside from strong emission lines and occasional absorption features,
the residual spectrum is nearly zero for almost all TTS.  A few TTS, 
such as the primary and secondary of DF Tau, display absorption features in the 
residual spectrum due to template mismatch.  Otherwise, the model
provides a good fit to spectra of all stars, including strong emission 
TTS such as the primary of XZ Tau.

Table 3 lists the veiling results for the medium resolution data.
We choose $\lambda_{ref}$ = 6100 \AA\ as a convenient continuum
band relatively free of strong absorption bands where our spectra
have good signal-to-noise.
The table includes the template, the veiling at 6100 \AA, $r_{6100}$,
defined in equation (4), the dereddened stellar flux at 6100 \AA,
and the dereddened veiling flux at 6100 \AA.  The veiling model
fails for stars without entries in any column. 

Figure 4 and Table 4 describe veiling results for the low resolution
spectra.  The format of the spectra are the same as in Fig.~3.  Most
residual spectra are flat over 5700--10000 \AA; some TTS have strong
emission lines from H$\alpha$, He~I, Na~I, [S II], and the Ca II
near-IR triplet.  A few TTS have upturns or downturns near the edges 
of their spectra.  These features are mostly caused by 
template mismatch (for example, they appear when using an M0 as a template for 
V999 Tau p, itself an M0.5 template), and by uncertain calibration
at long wavelengths.  These uncertainties have basically no effect on
the reddening, stellar luminosities, or effective temperatures.
The veiling parameters in Table 4 include the template, the veiling at 6100 \AA~and at 
8100 \AA, and the dereddened stellar flux at 6100 \AA.

\subsection{Physical Properties of T Tauri Stars}

Table 5 lists the derived physical properties of the TTS in our 
STIS sample.  The spectral types and $A_V$ in columns (2) and (3)
result from the veiling analysis of the low-resolution spectra. 
The reddening is the best $A_V$
derived from the veiling model for these spectra.
The formal error in $A_V$ is $\pm$ 0.2 -- 0.3 mag for most stars.
The spectral type is derived from the spectral types of the
templates for the low resolution and the medium resolution veiling
analysis.  Errors are $\pm$ 0.5 subclass for most stars;
the error is $\pm$ 1 subclass for XZ Tau p.  The scale of the spectral 
types is based on CaH spectral
types for nearby stars observed with the IRS \citep{ken01}. \citet{kir91}
use CaH to place these stars on the standard Boeshaar scale.  The other
parameters in this table are the stellar luminosity $L_{\star}$,
the accretion luminosity $L_{acc}$, the stellar mass $M_{\star}$,
the stellar age $\tau$, the stellar radius $R_{\star}$, the
accretion rate $\dot{M}$ of the disk onto the star, the infrared excess
and a quantity $\Delta$ log L$_{SEC}$ that measures how close the primary
and secondary are to being the same age.  The following paragraphs describe our
derivation of these parameters from our spectra and the veiling 
analysis.

Luminosity estimates are based on the \citet[][KH95 hereafter]{kenyon95} 
photometric compilation.  For each component of a close pair, our 
veiling analysis yields the R-band magnitude for the stellar photosphere 
and the veiling source.  These magnitudes represent the fraction 
of light passing through the STIS slit. To correct for any slit losses,
we assume that the total light from the pair equals the average R-band 
flux in KH95.  We then apportion the total R-band flux to each stellar
component and each veiling component of the pair.  For an adopted distance
of 140 pc for the Taurus-Auriga complex \citep{ken94a}, the stellar luminosity
of each component follows from a bolometric correction listed in KH95
derived for the stellar spectral type listed in Table~5.  

Estimates for the accretion luminosity rely on a bolometric correction
to convert the total flux in an observed bandpass to the total flux 
in the entire spectrum.  Because detailed physical models for the veiling
source are not available, a robust estimate for the bolometric correction
is not possible.  We follow \citet{har91}, \citet{val93}, \citet{gullbring98},
and others and adopt a simple slab model for the veiling emission.  The 
model specifies the veiling radiation by the temperature, the density, 
and the optical depth at a fixed wavelength \citep{kh87,har91}.  This model
accounts for many observed features of the veiling emission with a minimum 
of free parameters \citep[e.g.][]{ken94b,val93}.  

\citet{gullbring98} derive bolometric corrections for a range of slab models.
Their Fig.~9 shows that the ratio of the total flux to the flux in 
a 3200--5300 \AA~bandpass is $\sim$ 3--5 for temperatures $T \approx$
6000 K to 20000 K and optical depths $\tau_{3646}$ = 1--10 at a density 
$n_H = 10^{14}$ cm$^{-3}$.  They adopt a bolometric correction of 3.5
to convert their veiling observations to accretion luminosity.  Using 
the slab model described in \citet{har91}, we confirm the \citet{gullbring98} 
result for a modest range in hydrogen density, log $n_H$ = 13.5--14.5. 
For our 6000--6500 \AA~medium resolution passband, the typical bolometric
correction is $\sim$ 35.  Because the veiling model works well in the
6000--6500 \AA~region, we adopt this bandpass instead of the broader low
resolution bandpass.  For d = 140 pc, the accretion luminosity is then
\begin{equation}
L_{acc} \approx 2.14 \times 10^{10} ~ L_{\odot} ~ F_{veil,tot} ~ ,
\end{equation}
where $F_{veil,tot}$ is the flux from the veiling model integrated over
the medium resolution bandpass.

We follow WG01 (their Table~5) to convert the observed spectral types into
stellar effective temperatures. The effective temperature and
stellar luminosity give the stellar radius, and the stellar luminosity
and effective temperature place the star in the HR diagram, where masses and
ages result from pre-main-sequence evolutionary tracks. 

The accretion rates for TTS depend on the accretion geometry.
We adopt a magnetosphere model, where the stellar magnetic field truncates 
the disk a few stellar radii above the photosphere. The central 
star corotates with the inner disk; disk material then flows toward the 
star along magnetic field lines \citep[][1979b]{kon91,gho79a}.  Bright 
spots form where the accretion streams impact the star; these spots rotate 
with the stellar photosphere.  This geometry provides a natural explanation 
for the spectral energy distributions, optical photometric variations, and 
emission line profiles in many cTTS \citep{ber88,edw94,ken96}. 

The total infall energy per second from a magnetosphere model is
\begin{equation}
L_{infall} = \frac{G M_{\star} \dot{M}_{acc}}{R_{\star}} \left( 1 - \frac{R_{\star}}{R_{in}} \right) ~ ,
\end{equation}
where $G$ is the gravitational constant, $\dot{M}_{acc}$ is the accretion rate,
and $R_{in}$ is the inner radius of the disk.  We adopt $R_{in}$ = 
3 $R_{\star}$ \citep{ken96} and solve for the accretion rate:

\begin{equation}
\dot{M}_{acc} \approx 0.48 \times 10^{-7} ~ M_{\odot} ~ {\rm yr^{-1}} ~ 
\left( \frac{L_{infall}}{L_{\odot}} \right) ~ \left( \frac{R_{\star}}{R_{\odot}} \right) ~
\left( \frac{M_{\star}}{M_{\odot}} \right)^{-1}
\end{equation}
\citet{gullbring98} derive a similar expression for the accretion rate, though they 
assume that all of the accretion energy is radiated in the 2$\pi$ steradians away from the
star; that is, L$_{infall}$ = L$_{acc}$, where L$_{acc}$ is the radiated luminosity
from the hot spots.  Depending on the optical depth and altitude of the bright spot
above the photosphere, some of the accretion energy radiated toward 
the star will heat the stellar photosphere and some may help to drive an
outflow.  Here we assume that 50\% of the accretion energy is radiated
away from the star (L$_{infall}$ = 2L$_{acc}$).  The accretion rate is then

\begin{equation} 
\dot{M}_{acc} \approx 0.96 \times 10^{-7} ~ M_{\odot} ~ {\rm yr^{-1}} ~
\left( \frac{L_{acc}}{L_{\odot}} \right) ~ \left( \frac{R_{\star}}{R_{\odot}} \right) ~
\left( \frac{M_{\star}}{M_{\odot}} \right)^{-1}
\end{equation}

Infrared excesses for each star use the K and L colors compiled by
WG01. Because we have measured both the reddening and spectral type of each star from its
spectrum, we can correct the observed color for reddening and subtract the
photospheric colors to obtain a true infrared K$-$L color excess. 
Finally, the amount that the secondary must move in log L$_{\star}$ in the HR
diagrams of \citet{siess00} to make it the same age as the primary appears in
the last column of Table~5. Negative values imply that the secondary is younger than
the primary, and vice-versa for positive values.

\section{Disk and Accretion Properties of Close Binaries}

The main goal of this project is to determine how disk accretion behaves in
close binary systems. Our data are well-suited to compare primary disks
with those of their secondaries, because we have spectra of each component and
can therefore estimate the reddening, veiling, luminosities, masses, ages, mass accretion
rates throughout the sample.  In many cases infrared colors of each component also
exist in the literature. However, our sample is not
large enough to compare global star formation questions such as how quickly
disks evolve with age. Our sample deliberately avoids single stars, and so is not
representative of the stellar population as a whole.
For these reasons, in this paper we compare disk properties {\it within each pair} to try to gain 
some understanding about accretion in binaries.

\subsection{Mixed Pairs}

One of the most basic questions about young binaries is whether or not
a disk exists around one or both of the stars.
Although most binaries consist of two cTTs or two wTTs, about 15\%\ of young
binaries are `mixed pairs' that have a cTTs and a wTTs \citep{prato97,duchene99,wg01}.
The new HST spectra allow us to investigate this frequency among the closest pairs
by checking for disks around each star through
veiling, [O~I] emission, and strong H$\alpha$ emission. The spectra also reveal
the reddening and spectral types of each star individually, which we can combine
with existing photometry to detect even weak infrared excesses. Together, these four
disk signatures provide a thorough means to look for disks in the sample.
Because the medium-resolution spectra do not cover H$\alpha$, we cannot use the
width of the H$\alpha$ line to distinguish between cTTs and wTTs \citep[e.g.][]{bat96}.

Classification results appear in Table~7. In the Table, an `X' means that the
star has the disk signature, and a `$-$' means that it does not. We use the following
H$\alpha$ equivalent width criteria to determine
if a star is a cTTs -- EW(H$\alpha$) $>$ 5\AA\ for K stars, $>$ 10\AA\ for M0 -- M2,
and $>$ 25\AA\ for $>$M2.  With the exception of coolest stars, where
we have increased the EW from 20\AA\ to 25\AA\ to promote better consistency with the rest of our
data, these criteria are the same as those used by \cite{martin98}.
Stars are marked as cTTs if the veiling or [O~I] emission is $>$ 2$\sigma$.
For IR excesses, we use the K$-$L color of WG01, and 
subtract the reddened K$-$L photospheric color appropriate for the spectral type 
and reddening of each star (Table~5). If the resulting $\Delta$K$-$L $\ge$
0.20 magnitude, the object is a cTTs.  Of the 41 stars in Table~7, 28 (68\% of the sample)
have consistent classifications across all four disk signatures. 

Stars where one of the disk indicators disagrees with the other three provide useful
insights into the strengths and weaknesses of each of the criteria.
Nine stars (22\% of the sample) fall into this category. The least reliable indicator seems to be
[O~I], which gives a false negative (i.e., no [O~I] but all other disk signatures are present)
in 4 cases.  The lack of [O~I] for these cTTs may simply be signal to noise -- it is usually much easier to
see the [O~I] in the medium resolution spectrum than in the low resolution one because
[O~I] stands out more from the continuum in the medium resolution spectrum. The
M2.0 star V807 Tau~s, gives a false positive; it has strong [O~I] but no other disk signatures.
We have no explanation for this anomaly.  Similarly, the secondary of IS Tau, an M3.5, has weak [O~I]
but lacks other indications of a disk (the EW(H$\alpha$) of 24\AA\ falls just below our cutoff for 
a cTTs). 

The H$\alpha$, veiling, and IR-excess tests are each fairly reliable, but none are foolproof.
For example, the H$\alpha$ test gives
a false negative for the secondary of V955 Tau (6\AA , M2.5), which is a cTTs according to the [O~I]
emission and IR excess.  Had we used the H$\alpha$ cTTs criteria of \cite{martin98} there would be
three false positives, but we have avoided these by increasing the EW(H$\alpha$) cutoff for 
cTTs to 25\AA\ from 20\AA\ for the latest spectral types.
Veiling works about as well as H$\alpha$, giving false positives
for the secondaries of V927 Tau (r=0.09$\pm$0.03) and V999 Tau (r=0.29$\pm$0.05). The photosphere of
the latter is apparently anomalous.  In general,
veiling is easier to measure in later-type stars than in earlier-types because the temperature
differences between the accretion spots and the photospheres are higher when the photospheres are cool.
Finally, $\Delta$ K$-$L fails only in the primary of FQ Tau (M3.0), which shows no IR excess despite having
relatively strong veiling and [O~I] emission, as well as
an H$\alpha$ equivalent width of over 100\AA . Unlike veiling, IR excesses are less sensitive to the
presence of a disk when the photosphere is very red, so it is not too surprising that very red cTTs
sometimes lack excesses (see also the secondary of FO Tau below).

Four stars (10\% of the sample) in Table 7 show two
cTTs indicators and two wTTs indicators. The primary of V807 Tau is a K5 star with
a large IR excess and an H$\alpha$ equivalent width of $>$ 7\AA , so we classify this as
a cTTs. The lack of veiling is not particularly worrisome because the photosphere is relatively
blue, and there are at least four other cTTs without [O~I] (see above). The primary of
V955~Tau is another relatively blue star (K7) that has no detectable veiling or [O~I], but
satisfies the cTTs classification with 11\AA\ H$\alpha$ equivalent width and a large K$-$L
excess; we also classify this star as cTTs. As described above, its secondary has
strong [O~I] and a moderate infrared excess but lacks veiling and has weak H$\alpha$.
We classify this borderline case as a cTTs.  Finally, the secondary
of FO Tau must also be a cTTs; this M3.5 star has an H$\alpha$ equivalent width of 95\AA\ and a significant
amount of veiling. We attribute the lack of IR excess to its cool photosphere, and the lack of [O~I]
simply brings the total number of cTTs in that category up to seven. 

With these classifications, we have 12 cTTs pairs, 5 wTTs pairs, and 4 mixed pairs. The mixed pairs are
FQ Tau, FV Tau/c, IS Tau, and V807 Tau; the cTTs is the primary except in FV Tau/c. However, FV Tau/c
is not anomalous, as other examples of mixed pairs where the cTTs is the secondary exist in the literature
\citep{jaya01}.

This frequency of mixed pairs among the close binaries is essentially identical to that found
for wider binaries with larger samples \citep[][WG01]{duchene99}. 

\subsection{Mass Accretion Rates, Emission Lines, and IR Excesses}

Figure 5 shows the mass accretion rates of all pairs that have at least one
veiled component.  Ignoring the FV/FVc point, which plots the widely separated
primaries of these close binaries, the mass accretion rates of the primaries are correlated with
those of their secondaries. Hence, the disks around the closely separated systems that
comprise our sample are not independent of one another.
The significance of this correlation is very high -- a Spearman rank of 0.816 for 15 points
occurs with a probablility of less than $5\times 10^{-4}$ in uncorrelated data.
Surprisingly, though the mass accretion rates of the primaries 
exceed those of the secondaries in several cases, most of the sample points lie close to the
line of equal mass accretion rates.  In fact, 
when considered per unit mass (bottom panel in Fig.~5), secondaries appear somewhat more
active than primaries do, on average. 

The H$\alpha$ luminosity of all pairs in the sample (Fig.~6, top panel) also shows that activity levels
of the primaries and secondaries are correlated. Considering the sample as a whole is not particularly
useful however, because the paucity of mixed pairs implies a dearth of points in the lower right and upper
left portions of the graph, and this fact alone produces a correlation between the two variables.
A more interesting test is to compare the correlation among the cTTs pairs (solid squares in the figure). A
correlation does exist among cTTs, although there are only 12 points in this subsample. 
The H$\alpha$ correlation exists at a 99\%\ confidence level according to a Spearman rank test.
Most pairs lie to the right of the line in the top panel of Fig.~6; primaries
have larger intrinsic H$\alpha$ luminosities than secondaries do in the majority of pairs. 

When the H$\alpha$ luminosities are normalized by the stellar luminosities, the correlation among the cTTs
subsample vanishes, although it remains for the entire sample. Hence, the correlation in H$\alpha$
among the cTTs in Fig.~6 results from a correlation in stellar luminosities, which in turn arises 
because primaries and secondaries tend to be coeval (see section 5).

Our data are particularly well-suited to examine infrared excesses from colors because 
the extinction and the spectral types, both of which affect colors, are known for all the stars.
The K$-$L excesses in the bottom half of Fig.~6 also correlate for the primaries and
secondaries, but, as for the H$\alpha$ luminosities, this correlation is more interesting for the cTTs
subsample. A weak correlation between the K$-$L excesses of the primaries and their
secondaries, but this correlation rests entirely upon two points, XZ~Tau and FS~Tau. The correlation
exists at the 98\%\ confidence level, but without XZ~Tau and FS~Tau the confidence drops to $<$ 80\%.
K$-$L excesses arise in part from accretion in the disk, and correlate with veiling
\citep{hartigan90}; hence, a correlation of K$-$L colors within a pair
is likely to arise from the accretion rate correlations between primaries and
their secondaries. 

\subsection{Extinction}

Extinctions of primaries and their secondaries correlate (Fig.~7), as
expected for stars that lie along a nearly common sight line through a dark cloud.
The correlation in extinctions is very strong, with a confidence level of $>$ 99.99\%.
However, in six pairs (GH~Tau, UZ~Tau-W, UY~Aur, DD~Tau, V999~Tau, and FV~Tau/c),
the reddening of the secondary is significantly larger than that of the primary.
Only one of these, V999~Tau, is a wTTs pair; FV~Tau/c has a wTTs primary and cTTs
secondary, while the components of the other four pairs are all cTTs. As noted above,
the photosphere of V999~Tau~s appears anomalous, so it is possible that the extinction
may be more uncertain there.
Ignoring the FV/FVc point, which refers to the primaries of the FV~Tau and FV~Tau/c
systems and as such is a wide binary, there are only a couple of marginal cases of a primary that
appears more reddened than its secondary.

The tendency of some secondaries to have more extinction than their primaries has a 
fairly simple geometric interpretation for close binaries if the orbital plane of the binary
differs from that of the disk around the primary. In such systems the secondary will be
seen through the disk of the primary over part of its orbit.
If disks around primaries are, on average, more massive than those around secondaries,
then secondaries seen through the disks of their primaries will experience a
significant amount of additional extinction that would be less noticeable for a primary
seen through the disk of its secondary.  Because FV~Tau/c is a mixed pair with a wTTs primary,
additional local extinction around the secondary must either arise from the disk
of the secondary or from an extended envelope around the primary that lacks an
inner disk. More close pairs are needed to test if this trend of
more highly reddened secondaries is real.

\section{Ages, Masses, and Pre-Main-Sequence Tracks}

The binaries in our sample are plotted in an HR diagram together with the pre-main-sequence
evolutionary tracks of \citet{baraffe98}, \citet{siess00}, and \citet{palla99} in Figs.~8, 9,
and 10, respectively. Although most of the secondaries are less luminous than their primaries,
two pairs (XZ~Tau and FS~Tau) have more luminous secondaries which must be 
significantly younger than their primaries. The secondary of XZ~Tau dominates the mid-IR emission
from the pair \citep{krist97}, and both XZ~Tau and FS~Tau are surrounded by reflection nebulae. 
It is possible that the reflected light increases the apparent luminosities of the secondaries,
thereby decreasing their ages. The reflection nebulae indicate that the environments of these
binaries are very dusty, so it is conceivable that large grains in the immediate vicinity of
the primary produce a more grey reddening law, which would increase the extinction and
make the primary appear erroneously old. For this effect to equalize ages it would have
to act in the primary but not the secondary. There is no direct evidence to support this scenario.
Uncertainties of the stellar luminosities are dominated by systematic errors,
and are about 0.2 in log L.

As Fig.~11 shows, there is no clear correlation between the amplitude of 
$\Delta$ log L$_{SEC}$, the amount in log L that the secondary must shift to
be coeval with the primary, and the mass ratio. Secondaries tend to be younger
than primaries for both the Siess00 and Palla00 tracks (the BCAH98 tracks in Fig.~8
cannot be used because they do not extend to high enough masses or young enough ages).
In pairs such as DD Tau where the primary and secondary have the same spectral type,
and the tracks are nearly vertical in the HR diagram, the primary and secondary
will have the same mass but the primary will always appear younger because by our
definition the star we call the primary is the more luminous one at 7000\AA .
Hence, any statistics about age differences between primaries and secondaries must
eliminate these systems.
Ignoring such pairs, and leaving out the wide binary FV/FVc, the Palla00
tracks have 13 pairs where the secondary is younger and only 3 where it is older.
The corresponding numbers for the Siess00 tracks are 12 and 4, respectively.
This trend for the secondaries to be younger is unlikely to be caused by random
errors.  The chance that at least 13 out of 16 pairs in a coeval sample
have the secondary younger owing to random errors is only 1.1\% ;
the chance for at least 12 out of 16 pairs is 3.8\%.

WG01 noted that the age differences between components within a binary were significantly
less than those obtained by randomly pairing stars within their sample. This correlation
between the ages of primaries and secondaries is quite clear in Fig.~12, and exists at
the $3.5\times 10^{-3}$ level for the Palla00 tracks and at the $9\times 10^{-4}$ level
for the Siess00 tracks (Fig.~12 shows only the Siess00 tracks, because these are
the only ones with young enough isochrones for our sample).
The points in Fig.~11 also show the age correlation indirectly -- the standard
deviation of $\Delta$ log~L$_{SEC}$ is $\sim$ 0.28 for both sets of tracks, but should
be 0.51 if the stars in the sample were paired randomly. 

If, on average, secondaries formed some time after their primaries (for example
from instabilities in the primary's disk \citep{shu90}), then
the age differences should be largest on a log $\tau$ plot for the youngest
pairs because the fractional age difference declines as the pair ages.
Fig.~12 does not show this trend -- the two oldest primaries have the
largest age differences from their secondaries. 

One possible explanation for the younger ages of secondaries may be disk accretion \citep{hartmann97}.
If a 0.1 M$_\odot$ star were to accrete an additional 0.1 M$_\odot$ on a timescale short
compared to the age of the star, the new 0.2 M$_\odot$ star is in some sense
half as old as it once was. For this case of a mass-weighted age, the effect of accretion 
on age should depend upon $\dot{M}$/M (bottom panel of Fig.~5), where secondaries appear
somewhat more active than primaries, on average. However, the difference between primaries
and secondaries in Fig.~5 is less than a factor of two, and the effects 
of accretion on the location of a star in the HR diagram is a complex problem \citep[e.g.][]{tout99}.
There is no evidence that the secondaries continue to accrete matter after the primary
has stopped accreting -- if this were the case then most of the mixed pairs should have
a cTTs secondary and a wTTs primary, which they do not.
Although no set of pre-main sequence tracks can make all the pairs coeval, the
points in Fig.~12 would fall more closely on the line of equal ages if the 
isochrones were a little less steep in the HR diagram. Such behavior is indeed predicted by
some of the accretion models of \citet{tout99}. With the exception of the primary
of FS~Tau and possibly the primary of V955~Tau, all of the stars lie
on the convective portion of the tracks.

One advantage of studying close binaries is that they are sometimes surrounded by
rotating molecular disks in Keplerian orbits, so the combined mass of the system can
be estimated independently from the evolutionary tracks. The closest pairs also
show substantial orbital motion over a few years, which leads to a mass estimate for
the binary.  Dynamical masses from disk orbital motion have now been measured accurately
around a few T~Tauri stars \citep{duvert98,simon00,guill99}.
The best example is probably GG~Tau, which has a dynamical mass of 1.28 $\pm$ 0.07 M$_\odot$, 
somewhat larger than the 0.98 M$_\odot$ 
derived from the the location of the primary and secondary in
the HR diagram using either the \cite{siess00} tracks or the \cite{palla99} tracks.
If we take these numbers at face value, then an additional mass of 0.30 M$_\odot$ must be
in the system but not in the stars.  This extra mass is about 30\%\ of the stellar masses,
and seems large for disks, especially in a relatively unobscured system like GG~Tau.
A circumbinary disk also exists around UY~Aur with a dynamical mass of $\sim$ 1.2 M$_\odot$
\citep{duvert98}; the combined mass of the system using the Siess00 tracks is 0.94 M$_\odot$.
Following the assumptions described in the Notes for FS Tau leads to 1.2 M$_\odot$ in
that system, again larger than the combined mass of 0.89 M$_\odot$ in Table~5.

Dynamical masses of T Tauri systems continue to improve as better orbits are measured for
close binaries. The current data imply that discrepancies between the dynamical masses and
stellar masses are significant.  Either the dynamical masses are
too large, or the stellar masses too small; the latter could indicate
a problem with the evolutionary tracks or the effective temperature scales, while 
errors in dynamical masses could easily arise from uncertain distances.  For example,
a 10\%\ decrease in the distance to Taurus (to 126~pc) would decrease dynamical masses by $\sim$ 30\% ,
but would have little effect on the spectral masses because the pre-main-sequence tracks are nearly
vertical in the HR diagram for low mass stars, and changing the distance affects the luminosity
but not the effective temperature of the star.

\section{Summary}

This paper reports the results of a spectroscopic survey of 20 close (subarcsecond) binary
T Tauri stars in the Taurus-Auriga dark cloud done with STIS on the Hubble Space Telescope.
By obtaining spectra of each component it is possible to determine the reddening, spectral types,
and stellar luminosities of the primaries and secondaries in all the pairs.
This information suffices to place each star in the HR diagram, which determines its mass and age.
When combined with disk characteristics such as mass accretion rates measured from veiling emission,
and including emission line luminosities and infrared color excesses, it is possible to address many
of the outstanding questions regarding binary formation in systems where the stellar separations
are less than the extent of the circumstellar disks.  Such systems are not only interesting from
the standpoint of accretion dynamics, but also make up the most common mode of star formation in our
galaxy. 

Components of close binaries share common characteristics in a number of ways. Correlations in
mass accretion rates particularly stand out --
the best place to look for a rapidly accreting secondary
is next to a rapidly accreting primary. The actual values of the accretion rates are remarkably
similar between primaries and secondaries. Because secondaries are less luminous than their
primaries in most cases, it is often easier to see accretion signatures such as emission lines
and veiling around secondaries than it is around primaries. On average, secondaries actually
accrete more per unit mass than their primaries do.
Other accretion signatures such as H$\alpha$ emission and color excesses also correlate within
pairs, probably as a result of the correlation of these quantities with mass accretion rates.
Extinctions of primaries correlate well with those of their secondaries, as expected. 

We have examined four independent signatures of disks: H$\alpha$ emission, K$-$L color excesses,
veiling and [O~I] in each of our objects, and have devised rules each of these
signatures must satisfy to classify a star
as a cTTs. Most stars are clearly a cTTs or a wTTs, but several have at least one misleading
indicator, most often [O~I] although each of the four signatures is inaccurate in at least one case. 
Overall, the frequency of mixed pairs, where one star is a cTTs and one a wTTs, comprise about
the same fraction among close pairs as they do among wider pairs.

Several subtle trends have emerged from the data set. When extinctions differ within a pair it
is usually the secondary that is more heavily obscured. One can explain this result geometrically
by having a larger, more opaque disk around the primary than exists around the secondary.
Ages of primaries and secondaries also clearly correlate, and are equal to within the errors in
most cases. However, there is a clear tendency for secondaries to be a bit younger than their primaries.
Slightly flatter isochrones in the pre-main-sequence tracks, perhaps caused by ongoing accretion,
would remove this bias. However, pairs such as FS~Tau and XZ~Tau have more
luminous secondaries that will be very difficult to make coeval with their primaries for any set of tracks. 

Masses of the pairs inferred from their locations in the HR diagram are $\sim$ 30\% less than those
measured from rotating Keplerian disks or from the orbital motion within the binary. These 
discrepancies are most easily removed by reducing the distance to Taurus by about 10\%\ to
$\sim$ 126~pc, but can also be accomplished in other ways, such as altering the effective temperature scale
for cool photospheres or by generating a substantially different set of pre-main-sequence tracks.

\acknowledgements{This work has been supported under NASA/{\it HST}
grant GO-07310, from the Space Telescope Science Institute.}

\newpage

\centerline{Notes on Individual Pairs}

\noindent
DD Tau -- The primary and secondary are a little later than the
template, LkCa 7 s, at $\lambda > 8000$ \AA.  The veiling on the
medium resolution spectrum of DD Tau p is somewhat larger than that of
the low resolution spectrum.  Both spectra for DD Tau s yield the same veiling.
The reddening difference between the primary and secondary is significant.
The Ca II lines in the primary and the [O~I] lines
in the secondary are particularly strong. The secondary is the only star
in our sample with detectable [O~I]$\lambda$5577 emission. 

\noindent
DF Tau -- Residuals for the medium resolution spectra of both 
stars show a small wave, possibly resulting from a template mismatch.
The best template is V807 Tau s, which gives good residuals in the 
the low resolution spectra.  The primary has strong Paschen lines
that underlie the Ca II IR triplet.  The reddening is essentially the same
for the two stars.

\noindent
FO Tau -- Residuals for the medium resolution spectra of both 
stars show a low-amplitude wave, like those of DF~Tau only weaker. 
Residuals in the low resolution spectra are flat out to 9000 \AA. 
LkCa 7 s is the best template for both primary and secondary. 
FO Tau p has Paschen lines in emission. Although
the secondary has no [O~I] emission or IR excess, its H$\alpha$ equivalent width
is 95\AA , which clearly marks it as a cTTs.

\noindent
FQ Tau (Haro 6-3) -- The low- and medium-resolution spectra of both stars yield flat residuals
and consistent veiling results where they overlap in wavelength.  The secondary has 
an H$\alpha$ equivalent width of 23\AA , but this star is an M3.5 and has no other
accretion signatures so we classify it as a wTTs. The secondary is brighter than the
primary longward of about 7000\AA .
FQ Tau p has strong H$\alpha$, He I, and Ca II triplet lines, and is veiled, but
has no IR excess.  

\noindent 
FS Tau (Haro 6-5A) --
Forbidden lines are strong in this pair, which is surrounded by a reflection nebula
\citet{krist98}.  The red doublet line of [S II] $\lambda$6720 has
an equivalent width of 4.9 $\pm$ 0.4 \AA\ in the primary.
H$\alpha$ and [O~I] emission fill the space between the pair in the longslit data,
and extend brightly along the slit
for an additional 0.3 arcseconds past the primary and some 0.2 arcseconds beyond the
secondary. The [S~II] emission is also extended, and
H$\alpha$ continues weakly out to $\sim$ 1.7 arcseconds away from the primary.
A second system, FS~Tau~B located some 20$^{\prime\prime}$ west of FS~Tau, also drives a
highly collimated jet \citep{woitas02}.
The primary is too faint in the medium resolution spectrum 
to derive a reliable veiling, but the low resolution spectrum
yields a good veiling measurement that is relatively
insensitive to the reddening. The medium resolution spectrum of FS Tau s
has low S/N, but yields the same veiling as the low resolution spectrum of this star,
which is very red.  Binning the medium resolution spectrum to increase the S/N
yields a similar result.  The PA of the system has increased by $\sim$ 10 degrees from that
measured by \citet{krist98} from HST images taken about 5 years earlier. If the orbit is
circular and viewed pole-on, the period of 180 years implies a mass of 1.20 M$_\odot$
for d = 140~pc.

\noindent
FV Tau (Haro 6-8) -- Despite a large reddening, spectra of the primary yield
good veiling measurements.   The secondary is a continuum T~Tauri star that
shows no photospheric features and has strong H$\alpha$ and Ca II triplet emission.
The $\Delta$ K$-$L value in Table~5 assumes an intrinsic K$-$L = 0.16, corresponding to
an M2.0 star, and the same reddening as the primary. 
Regardless of the spectral type and reddening assumed, the star is clearly a cTTs.
We obtain an approximate mass accretion rate by taking the star to lie in the middle
of the sample of stars we observed, with mass 0.40 M$_\odot$ and radius 1.5 R$_\odot$. 

\noindent
FV Tau/c -- The primary star has no veiling on either spectrum. 
The secondary is very red and has very little flux $<$ 7000\AA , so the reddening
is difficult to measure, though is definitely greater than that of the
primary. The values for this star in Table~5 come from considering a range of A$_V$, and
choosing the value that makes the veiling reasonably flat. Visual extinctions less than 5.5
or greater than 8 lead to non-physical veiling solutions. 
The medium resolution spectrum for FV Tau/c s is too noisy to
derive a reliable veiling measurement.  The extremely large H$\alpha$ and
[O~I] 6300 equivalent widths are caused by the very weak continuum at those wavelengths. 
For $\lambda > 6000$ \AA, the veiling model fits the low resolution spectrum well. 
The pair is located about 12 arcseconds away from FV~Tau.

\noindent 
GG Tau -- The secondary was only partially contained within the slit. The
spectrum of the secondary was multiplied by a factor of 2.3 to compensate
for this effect, which brings the flux ratio of the spectra
of the two stars in agreement with the flux ratio of $\sim$ 3.0 measured from
the acquisition image. 
Residuals in the low resolution models for both stars have a small wave.
H$\alpha$ is saturated in the primary, which shows Ca II emission lines
as well as Paschen lines. The dereddened fluxes of the brightest Paschen lines
are about one-third that of the Ca II lines.

\noindent
GH Tau (Haro 6-20)-- Aside from a difference of 0.5 in the reddening, the spectra 
of these two stars are essentially identical at both low and medium resolution except for
[O~I] emission lines that occur in the secondary but not in the primary.  
On the medium resolution spectra, both stars have somewhat stronger absorption lines
than the templates do.  This result does not depend on the choice of the 
template and is not caused by template mismatch in spectral type.  
At low resolution, both stars have weaker TiO bands than exists in the templates.
If one star is used as a template for the other, the residuals show no
photospheric features.
The reddening difference between the primary and secondary is significant --
making the extinctions of the two stars equal introduces a negative veiling
in one or both components.

\noindent
HBC 358 (NTTS 040047) -- This system is a hierarchical triple. The secondary
consists of two components that differ by only 0.01 magnitude in the acquisition
image, and are each 0.28 magnitude fainter than the primary. The brighter of secondary
components and the primary were centered on the slit. About 1/3 of the
light from the fainter component of the secondary was included
within the slit, and coincides spatially with the spectrum from the brighter
component of the secondary. To correct for this effect, the extracted spectrum of
the secondary was multiplied by a factor of 0.75. Hence, the spectrum of the secondary
is that of the brighter component only.

\noindent
HBC 360 (NTTS 040142) -- HBC 360 is wide binary with a separation of about 7.1 arcseconds
and two nearly equal components. Only the primary was observed with HST, but the secondary
is also known to be a wTTs \citep{hartigan94}.

\noindent
IS Tau (Haro 6-23) -- Low and medium resolution spectra give consistent veilings for
both primary and secondary where the wavelengths overlap.  Both stars are have visual
extinctions of about 3.5 magnitudes.
IS Tau s has a somewhat later spectral type than the template, but choosing a later
template produces a poorer fit. The slight rise in the residual of the
low resolution model for IS Tau p is caused by a combination of noise in
the spectrum and spectral type mismatch in the template.

\noindent
Lk Ca 7 (V1070 Tau) -- Both the primary (K7) and secondary (M3.5) of this subarcsecond weak-lined
T~Tauri binary are useful as spectroscopic templates.

\noindent
UY Aur -- UY Aur s has bright Ca triplet emission. The template provides a good match to
UY Aur s on both spectra. Residuals for the medium resolution spectrum 
of UY Aur p have a small wave.  The reddening difference between the stars
is significant. The secondary of UY Aur has been known to have variable reddening since the
early observations of \cite{joy44} \citep[see also][]{herbst95}.

\noindent
UZ Tau/W -- Given the errors, both stars have no veiling in the low-resolution spectrum, but
veiling at the 2 -- 3 sigma level is present for both stars in the medium-resolution spectra. 
Both stars have strong H$\alpha$ emission and significant K$-$L excess, and [O~I] is also
bright in the secondary, so this is a cTTs pair with unusually weak veiling.
The templates provide an excellent match for the medium resolution spectra of both stars. Residuals
for the low resolution spectra have a wavy pattern at long wavelengths.
Changing the template makes this pattern worse. About 3.5 arcseconds to the east lies
UZ Tau~E, which \cite{mathieu96} report as a spectroscopic binary.
The secondary is significantly more reddened than the primary.

\noindent
V807 Tau -- The primary is a K5 with an H$\alpha$ equivalent width of 7.4\AA\ a large IR excess,
and so is a cTTs even though it is not veiled (and, in fact, used as a template). The secondary,
an M2.0, is also a good spectroscopic template. The secondary has relatively strong [O~I]
emission, but lacks other cTTs signatures.

\noindent
V927 Tau (LkH$\alpha$ 331) -- The low resolution spectrum of the primary is a little 
later than the template. 

\noindent 
V955 Tau (LkH$\alpha$ 332) -- The templates provide good matches to the spectra of both stars,
though the low resolution spectrum of V955 Tau s is slightly later than the template. 

\noindent
V999 Tau (LkH$\alpha$ 332/G2) -- The template for the primary star does not match as well as most of the rest
of the stars in the sample. The secondary spectrum yields a good veiling measurement at low resolution,
but shows no other cTTs signatures. 

\noindent
V1026 Tau (Haro 6-28) -- Low and medium-resolution spectra for
both stars give consistent veiling where the wavelengths overlap, and 
the templates provide a very good match to the absorption lines.

\noindent
XZ Tau - The veiling measurement for the primary is large. 
XZ Tau p has many emission lines.  The Ca II triplet is particularly strong here, with each line
nearly as bright as H$\alpha$. The Paschen series is visible from Pa7 near the red limit
of the spectra through Pa14, above which the series becomes weaker and blends with the Ca II lines.
The dereddened luminosities for the brighter Paschen lines are about $1.3\times 10^{-4}$ L$_\odot$ apiece. 
The reddening seems reliable -- larger values make the low resolution spectrum too blue.  Residuals for
the low resolution spectrum of XZ Tau s are relatively large at the longest wavelengths.

\clearpage
\begin{center}
\begin{deluxetable}{lrrr}
\singlespace
\tablenum{1}
\tablewidth{0pt}
\tablecolumns{4}
\tabcolsep = 0.28in
\parindent=0em
\tablecaption{Position Angles and Separations$^a$}
\tablehead{
\colhead{Name} & \colhead{$^b$PA} &  \colhead{$^c$Sep} & \colhead{Date}}
\startdata
DD Tau        &182.1 (1.0)& 568 (14)&1999.68\\
DF Tau        &273.4 (6.1)&  99 (14)&1998.90\\
FO Tau        &203.8 (4.1)& 146 (14)&1999.79\\
FQ Tau        & 67.4 (0.7)& 763 (14)&1998.93\\
FS Tau        & 93.7 (2.5)& 242 (14)&2000.95\\
FV Tau        &273.0 (0.8)& 711 (14)&2000.92\\
FV Tau/c$^h$  &293.6 (0.8)& 701 (14)&1998.92\\
GG Tau        &348.6 (2.4)& 248 (14)&2001.06\\
GH Tau        &293.7 (2.0)& 304 (14)&2001.06\\
HBC 358$^f$   & 47.5 (0.4)&1555 (14)&1998.92\\
HBC 358s$^g$  &337.3 (4.1)& 149 (14)&1998.92\\
IS Tau        &104.1 (2.8)& 216 (14)&2000.92\\
LkCa 7        & 25.2 (0.6)&1035 (14)&1999.07\\
UY Aur        &228.1 (0.7)& 892 (14)&1998.98\\
UZ Tau W      &  3.6 (1.6)& 369 (14)&1999.84$^e$\\
UZ Tau E$^d$  & 93.2 (0.2)&3551 (14)&1999.84\\
V807 Tau      &322.9 (2.0)& 297 (14)&1998.92\\
V927 Tau      &287.8 (2.1)& 280 (14)&1999.77\\ 
V955 Tau      &206.1 (1.7)& 339 (14)&1998.98\\ 
V999 Tau      &242.0 (3.1)& 226 (14)&1998.95\\ 
V1026 Tau     &245.9 (0.8)& 658 (14)&1998.92\\ 
XZ Tau        &142.6 (2.0)& 299 (14)&2000.92\\
\enddata
\tablenotetext{a} {As determined from the STIS acquisition images.}
\tablenotetext{b} {Position angle of the secondary relative to the primary (and errors) in degrees.}
\tablenotetext{c} {Separations (and errors) in milliarcseconds.}
\tablenotetext{d} {Separation between UZ Tau W pri and UZ Tau E}
\tablenotetext{e} {Spectra of UZ Tau W were taken 2000.68 UT, about a year after the
image used to measure the separations.}
\tablenotetext{f} {Separation between HBC 358 pri and HBC 358 sec-A.}
\tablenotetext{g} {Separation between HBC 358 sec-A and HBC 358 sec-B.}
\tablenotetext{h} {This binary is located about 12 arcseconds from the FV Tau pair.}
\end{deluxetable}
\end{center}

\clearpage
\begin{center}
\begin{deluxetable}{l r c c c c c c c}
\tablecolumns{9}
\tablewidth{0 pt}
\tablenum{2}
\tablecaption{Observed Equivalent Widths of Emission Lines$^a$}
\tablehead{ 
\colhead{} & \colhead{H$\alpha$} & \colhead{[O I]} & \colhead{[O I]} &
\colhead{Ca II} & \colhead{Ca II} & \colhead{Ca II} & \colhead{Pa9}\\
\noalign{\vskip -7pt}
\colhead{Name$^b$} & \colhead{6563} & \colhead{5577} & \colhead{6300} &
\colhead{8498} & \colhead{8542} & \colhead{8662} & \colhead{9229} }
\startdata
DD Tau p   &    206\ (10)&$<$ 0.4 &    7.7\ (0.2) &7.8\ (0.4)&5.9\ (0.4)&6.1\ (0.4)&$<$ 0.5\\
DD Tau s   &   635\ (30)&9.1 (1.7)&     19\ (01)  &$<$ 0.8   &2.7\ (1.2)&1.6\ (1.0)&$<$ 0.6\\
DF Tau p   & $^c$55\ (8)&$<$ 0.15 &$^d$1.8\ (0.2) &2.0\ (0.2)&3.0\ (0.2)&1.8\ (0.2)&2.4\ (0.2)\\
DF Tau s   & $^c$52\ (8)&$<$ 0.2  &$^d$1.7\ (0.2) &2.1\ (0.3)&1.6\ (0.2)&1.6\ (0.2)&$<$ 0.4   \\
FO Tau p   &   137\ (10)&$<$ 0.2  &    0.5\ (0.1) &2.6\ (0.2)&2.5\ (0.2)&2.8\ (0.3)&1.3\ (0.2)\\
FO Tau s   &    95\ (10)& $<$ 0.5 &    $<$ 0.1    & $<$ 0.5  & $<$ 0.5  & $<$ 0.5  &$<$ 0.3  \\
FQ Tau p   &   110\  (5)& $<$ 0.5 &   0.9\ (0.1)  &5.9\ (0.5)&6.3\ (0.4)&4.0\ (0.4)&$<$ 0.5 \\
FQ Tau s   &    23\  (4)& $<$ 0.9 &    $<$ 0.1    & $<$ 0.5  & $<$ 0.5  & $<$ 0.5  &$<$ 0.6  \\
FS Tau p   &    71\  (5)& $<$ 2.5 &$^e$21\ (3)    &3.7\ (0.3)&4.9\ (0.3)&5.4\ (0.3)&1.1\ (0.2)\\
FS Tau s   &    33\  (3)& $<$ 8.0 &    13\ (2)    &$<$ 0.8   &$<$ 0.5   &$<$ 0.8   &$<$ 0.3   \\
FV Tau p   &    15\  (2)& $<$ 0.7 &   0.3\ (0.1)  &2.7\ (0.8)&0.8\ (0.8)&2.0\ (0.8)&$<$ 0.9   \\
FV Tau s   &    63\  (5)& $<$ 1.5 &   1.6\ (0.3)  &8.8\ (0.4)&9.7\ (0.5)&6.7\ (0.3)&$<$ 0.9   \\
FV Tau/c p &    21\  (2)& $<$ 0.6 &  $<$ 0.1      &$<$ 0.4   &$<$ 0.4   &$<$ 0.4   &$<$ 0.4 \\
FV Tau/c s &   800\ (90)& $<$ 12  &    67\ (20)   &$<$ 1.2   &$<$ 1.0   &$<$ 0.6   &$<$ 1.4   \\
GG Tau p   &$^c$42\  (2)& $<$ 0.7 &  $<$ 0.2      &2.9\ (0.4)&2.8\ (0.4)&2.2\ (0.3)&1.5\ (0.3)\\
GG Tau s   &    21\  (4)& $<$ 0.4 &   0.5\ (0.1)  & $<$ 0.5  & $<$ 0.5  & $<$ 0.5  &$<$ 0.5\\
GH Tau p   &    10\  (1)& $<$ 0.6 &  $<$ 0.2      & $<$ 0.3  & $<$ 0.3  & $<$ 0.3  &$<$ 0.3\\
GH Tau s   &    10\  (1)& $<$ 0.8 &  0.9\ (0.1)   & $<$ 0.6  & $<$ 0.6  & $<$ 0.7  &$<$ 0.7\\
HBC 358 p  &  4.5\ (0.4)& $<$ 1.1 &    $<$ 0.1    & $<$ 1.1  & $<$ 1.1  & $<$ 1.1  &$<$ 1.1   \\
HBC 358 s  &  6.9\ (1.0)& $<$ 1.1 &    $<$ 0.1    & $<$ 1.1  & $<$ 1.1  & $<$ 1.1  &$<$ 1.1   \\
HBC 360 p  &  4.6\ (0.4)& $<$ 1.1 &    $<$ 0.2    & $<$ 1.1  & $<$ 1.1  & $<$ 1.1  &$<$ 1.1 \\
IS Tau p   &  9.5\ (0.4)& $<$ 0.2 & 0.5 (0.1)     & $<$ 0.9  & $<$ 0.9  & $<$ 0.8  &$<$ 0.8 \\
IS Tau s   &    24\ (3) & $<$ 1.8 & 0.5 (0.2)     & $<$ 0.8  & $<$ 0.8  & $<$ 0.8  &$<$ 0.8\\
LkCa 7 p   &  1.4\ (0.2)& $<$ 0.5 &    $<$ 0.1    & $<$ 0.5  & $<$ 0.5  & $<$ 0.5  &$<$ 0.5\\
LkCa 7 s   &  2.2\ (1.0)& $<$ 0.5 &    $<$ 0.1    & $<$ 0.8  & $<$ 0.8  & $<$ 0.8  &$<$ 0.8\\
UY Aur p   &    36\ (1) & $<$ 0.5 &$^d$1.1\ (0.1) & $<$ 0.2  & $<$ 0.2  & $<$ 0.2  &$<$ 0.2\\
UY Aur s   &    59\ (4)& $<$ 0.6 &    3.4\ (0.2) &0.5\ (0.6)&10.2\ (0.6)&8.5\ (0.6)&$<$ 0.6\\
UZ Tau W p &    54\  (8)& $<$ 0.3 &   $<$ 0.1     & $<$ 0.5  & $<$ 0.5  & $<$ 0.5  &$<$ 0.5 \\
UZ Tau W s &    97\  (8)& $<$ 1.5 &   1.5 (0.1)   & $<$ 0.5  & $<$ 0.5  & $<$ 0.5  &$<$ 0.5 \\
V807 Tau p &$^c$7.4\ (0.5)& $<$ 0.4 &  $<$ 0.1    & $<$ 0.6  & $<$ 0.6  & $<$ 0.7  &$<$ 0.7 \\
V807 Tau s & 3.4\ (0.4)& $<$ 0.5 &   0.8 (0.1)   & $<$ 0.6  & $<$ 0.6  & $<$ 0.7  &$<$ 0.7\\
V927 Tau p & 5.6\ (0.8) & $<$ 0.8 &    $<$ 0.05    & $<$ 0.5  & $<$ 0.5  & $<$ 0.5  &$<$ 0.5\\
V927 Tau s & 2.5\ (1.8) & $<$ 0.8 &    $<$ 0.4    & $<$ 1.5  & $<$ 1.5  & $<$ 1.5  &$<$ 1.5\\
V955 Tau p &10.9\ (0.6) & $<$ 0.5 &    $<$ 0.2     & $<$ 0.4  & $<$ 0.4  & $<$ 0.4  &$<$ 0.4\\
V955 Tau s & 6.2\ (0.6) & $<$ 0.6 &   1.1 (0.1)    & $<$ 0.7  & $<$ 0.7  & $<$ 0.7  &$<$ 0.7\\
V999 Tau p &  1.6\ (0.3)& $<$ 0.6 &    $<$ 0.1    & $<$ 0.6  & $<$ 0.6  & $<$ 0.6  &$<$ 0.6\\
V999 Tau s &  4.5\ (0.4)& $<$ 1.5 &    $<$ 0.1    & $<$ 0.7  & $<$ 0.7  & $<$ 0.7  &$<$ 0.7\\
V1026 Tau p &   57\  (6)& $<$ 0.4 &  $<$ 0.1      & $<$ 0.3  & $<$ 0.3  & $<$ 0.4  &0.6\ (0.2)\\
V1026 Tau s &  124\ (10)& $<$ 2.0 & 5.6\ (0.5)    &$<$ 1.1   &$<$ 1.1   &$<$ 1.1   &$<$ 1.1   \\
XZ Tau p   &    77\  (6)& $<$ 0.4 &   2.8\ (0.1)  & 36\ (2)  & 36\ (1)  & 37\ (2)  &2.3\ (0.2)\\
XZ Tau s   &    42\  (5)& $<$ 0.6 &  11.5\ (0.5)  & $<$ 1.2  & $<$ 1.2  & $<$ 1.2  &$<$ 1.2 \\
\enddata
\tablenotetext{a} {Equivalent widths of emission lines in \AA , as measured from the observed spectra without
any modeling of photospheric absorption beneath the emission line profiles. Uncertainties are listed in parentheses.}
\tablenotetext{b} {Primaries are listed as `p' and secondaries as `s', where the
primary is defined as the earlier spectral type when the spectral types differ,
and as the brighter star at 7000\AA\ when the spectral types are the same.}
\tablenotetext{c} {H$\alpha$ is saturated so the number listed is a
lower limit to the equivalent width.}
\tablenotetext{d} {[O I] 6300\AA\ line has resolved velocity structure.}
\tablenotetext{e} {[O I] 6300\AA\ appears extended spatially.}
\end{deluxetable}
\end{center}
\clearpage
\begin{center}
\begin{deluxetable}{l c c c c}
\singlespace
\small
\tablewidth{0 pt}
\tablenum{3}
\tablecaption{Veiling Results from Medium Resolution HST Spectra$^a$}
\tablehead{
\colhead{Name} &
\colhead{Template} &
\colhead{$r_{6100}$} &
\colhead{$^b$log F$^{\prime}_{\star, 6100}$} &
\colhead{$^c$F$^{\prime}_{V,6100}$}}
\startdata
DD Tau p & LkCa 7 s    & $+0.84\ (0.04)$ & $-14.15\ (0.03)$ & $-14.23\ (0.02)$ \\
DD Tau s & LkCa 7 s    & $+0.53\ (0.04)$ & $-14.32\ (0.03)$ & $-14.59\ (0.03)$ \\
DF Tau p & V807 Tau s  & $+0.86\ (0.05)$ & $-13.60\ (0.02)$ & $-13.66\ (0.02)$ \\
DF Tau s & V807 Tau s  & $+0.52\ (0.05)$ & $-13.50\ (0.02)$ & $-13.78\ (0.02)$ \\
FO Tau p & LkCa 7 s    & $+0.45\ (0.05)$ & $-14.18\ (0.03)$ & $-14.52\ (0.03)$ \\
FO Tau s & LkCa 7 s    & $+0.32\ (0.04)$ & $-14.21\ (0.03)$ & $-14.72\ (0.02)$ \\
FQ Tau p & LkCa 7 s    & $+0.58\ (0.06)$ & $-14.40\ (0.03)$ & $-14.64\ (0.04)$ \\
FQ Tau s & LkCa 7 s    & $+0.03\ (0.04)$ & $-14.28\ (0.03)$ & $<\ -15.68$      \\
FS Tau p & \nodata     & \nodata         & \nodata          & \nodata          \\
FS Tau s & LkCa 7 s    & $+0.21\ (0.11)$ & $-14.47\ (0.07)$ & $-15.14\ (0.09)$ \\
FV Tau p & V807 Tau p  & $+0.25\ (0.02)$ & $-12.96\ (0.03)$ & $-13.57\ (0.04)$ \\
FV Tau s & \nodata     & \nodata         & \nodata          & \nodata          \\
FV Tau/c p &V807 Tau s & $-0.01\ (0.03)$ & $-14.06\ (0.03)$ & $<\ -15.58$      \\
FV Tau/c s &\nodata    & \nodata         & \nodata          & \nodata          \\
GG Tau p & LkCa 7 p    & $+0.22\ (0.02)$ & $-13.31\ (0.03)$ & $-13.97\ (0.04)$ \\
GG Tau s & V807 Tau s  & $+0.49\ (0.02)$ & $-13.84\ (0.03)$ & $-14.14\ (0.04)$ \\
GH Tau p & V955 Tau s  & $+0.13\ (0.02)$ & $-13.82\ (0.03)$ & $-14.71\ (0.04)$ \\
GH Tau s & V955 Tau s  & $+0.13\ (0.03)$ & $-13.82\ (0.03)$ & $-14.70\ (0.04)$ \\
HBC 358 p & HBC 358 s  & $+0.06\ (0.04)$ & $-14.61\ (0.02)$ & $<\ -15.83$      \\
HBC 358 s & HBC 358 p  & $-0.06\ (0.04)$ & $-14.64\ (0.02)$ & $<\ -15.86$      \\
HBC 360 p & HBC 358 p  & $+0.03\ (0.04)$ & $-14.35\ (0.02)$ & $<\ -15.75$      \\
IS Tau p & LkCa 7 p    & $+0.06\ (0.03)$ & $-13.43\ (0.05)$ & $-14.63\ (0.04)$ \\
IS Tau s & LkCa 7 s    & $+0.14\ (0.09)$ & $-14.42\ (0.07)$ & $<\ -15.27$      \\
UY Aur p & LkCa 7 p    & $+0.27\ (0.03)$ & $-13.43\ (0.02)$ & $-14.00\ (0.02)$ \\
UY Aur s & V807 Tau s  & $+0.67\ (0.03)$ & $-14.06\ (0.03)$ & $-14.24\ (0.03)$ \\
UZ Tau W p & V807 Tau s& $+0.06\ (0.03)$ & $-13.76\ (0.04)$ & $-14.99\ (0.06)$ \\
UZ Tau W s & V807 Tau s& $+0.04\ (0.02)$ & $-13.69\ (0.04)$ & $-15.14\ (0.06)$ \\
V927 Tau p & LkCa 7 s  & $+0.01\ (0.04)$ & $-14.08\ (0.02)$ & $<\ -15.48$      \\
V927 Tau s & LkCa 7 s  & $+0.09\ (0.03)$ & $-14.68\ (0.03)$ & $-15.72\ (0.10)$ \\
V955 Tau p & V807 Tau p& $-0.05\ (0.03)$ & $-13.19\ (0.01)$ & $<\ -14.49$      \\
V955 Tau s & V807 Tau s& $+0.03\ (0.03)$ & $-14.03\ (0.03)$ & $-15.60\ (0.10)$ \\
V999 Tau s & V955 Tau s& $+0.29\ (0.05)$ & $-14.13\ (0.03)$ & $-14.66\ (0.04)$ \\
V1026 Tau p &V807 Tau s& $+0.68\ (0.04)$ & $-14.30\ (0.03)$ & $-14.47\ (0.03)$ \\
V1026 Tau s & HBC 358 s& $+0.23\ (0.05)$ & $-15.10\ (0.04)$ & $-15.74\ (0.06)$ \\
XZ Tau p & V807 Tau s  & $+4.12\ (0.30)$ & $-14.13\ (0.15)$ & $-13.51\ (0.25)$ \\
XZ Tau s & HBC 358 s   & $+0.35\ (0.02)$ & $-13.90\ (0.03)$ & $-14.35\ (0.03)$ \\
\enddata
\tablenotetext{a} {Uncertainties in the last two digits appear in parentheses.}
\tablenotetext{b} {Dereddened stellar flux at 6100\AA , in units erg$\,$cm$^{-2}$s$^{-1}$\AA$^{-1}$}
\tablenotetext{c} {Dereddened veiling flux at 6100\AA , in units erg$\,$cm$^{-2}$s$^{-1}$\AA$^{-1}$}
\end{deluxetable}
\end{center}
   
\clearpage
\begin{center}
\begin{deluxetable}{l c c c c}
\singlespace
\small
\tablewidth{0 pt}
\tablenum{4}
\tablecaption{Veiling Results from Low Resolution HST Spectra$^a$}
\tablehead{
\colhead{TTS} & \colhead{Template} &
\colhead{$r_{6100}$} & \colhead{$r_{8115}$} &
\colhead{$^b$log $F_{\star (6100A)}$}}
\startdata
DD Tau p   & LkCa 7 s  & $+0.52\ (05)$ & $+0.10\ (02)$ & $-14.03\ (04)$ \\
DD Tau s   & LkCa 7 s  & $+0.54\ (05)$ & $+0.08\ (03)$ & $-14.24\ (04)$ \\
DF Tau p   & V807 Tau s& $+0.70\ (06)$ & $+0.29\ (04)$ & $-13.53\ (04)$ \\
DF Tau s   & V807 Tau s& $+0.61\ (05)$ & $+0.28\ (04)$ & $-13.47\ (04)$ \\
FO Tau p   & LkCa 7 s  & $+0.39\ (05)$ & $+0.19\ (04)$ & $-14.15\ (05)$ \\
FO Tau s   & LkCa 7 s  & $+0.26\ (03)$ & $+0.14\ (03)$ & $-14.15\ (05)$ \\
FQ Tau p   & LkCa 7 s  & $+0.54\ (07)$ & $+0.07\ (05)$ & $-14.31\ (04)$ \\
FQ Tau s   & LkCa 7 s  & $+0.07\ (05)$ & $+0.01\ (05)$ & $-14.27\ (04)$ \\
FS Tau p   & LkCa 7 p  & $+0.40\ (10)$ & $+0.34\ (11)$ & $-13.99\ (10)$ \\
FS Tau s   & LkCa 7 s  & $+0.08\ (09)$ & $+0.06\ (08)$ & $-14.40\ (11)$ \\
FV Tau p   & V807 Tau p& $+0.30\ (04)$ & $+0.42\ (04)$ & $-12.97\ (03)$ \\
FV Tau s   & \nodata   & $>$ 5         & \nodata       & \nodata        \\
FV Tau/c p & V807 Tau s& $-0.10\ (06)$ & $+0.03\ (04)$ & $-13.97\ (02)$ \\
FV Tau/c s & LkCa 7 s  & $+1.15\ (09)$ & $+0.24\ (06)$ & $-14.61\ (10)$ \\
GG Tau p   & LkCa 7 p  & $+0.59\ (05)$ & $+0.56\ (04)$ & $-13.41\ (03)$ \\
GG Tau s   & V807 Tau s& $+0.67\ (05)$ & $+0.58\ (04)$ & $-13.89\ (03)$ \\
GH Tau p   & V807 Tau s& $+0.18\ (05)$ & $+0.13\ (04)$ & $-13.82\ (03)$ \\
GH Tau s   & V807 Tau s& $+0.20\ (05)$ & $+0.19\ (04)$ & $-13.83\ (03)$ \\
HBC 358 p  & HBC 360 p & $+0.04\ (04)$ & $+0.00\ (03)$ & $-14.27\ (03)$ \\
HBC 358 s  & LkCa 7 s  & $-0.02\ (03)$ & $+0.02\ (03)$ & $-14.64\ (03)$ \\
HBC 360 p  & HBC 358 p & $-0.01\ (04)$ & $+0.03\ (03)$ & $-14.48\ (03)$ \\
IS Tau p   & LkCa 7 p  & $+0.02\ (05)$ & $+0.01\ (05)$ & $-13.36\ (06)$ \\
IS Tau s   & LkCa 7 s  & $+0.03\ (04)$ & $+0.02\ (04)$ & $-14.28\ (02)$ \\
UY Aur p   & LkCa 7 p  & $+0.47\ (06)$ & $+0.50\ (05)$ & $-13.46\ (04)$ \\
UY Aur s   & V807 Tau s& $+0.60\ (04)$ & $+0.47\ (04)$ & $-13.99\ (05)$ \\
UZ Tau/W p & V807 Tau s& $+0.04\ (03)$ & $+0.06\ (04)$ & $-13.73\ (04)$ \\
UZ Tau/W s & V807 Tau s& $-0.04\ (03)$ & $-0.10\ (04)$ & $-13.60\ (04)$ \\
V927 Tau p & LkCa 7 s  & $+0.01\ (04)$ & $+0.00\ (03)$ & $-13.97\ (04)$ \\
V927 Tau s & LkCa 7 s  & $+0.00\ (03)$ & $+0.03\ (03)$ & $-14.52\ (03)$ \\
V955 Tau p & V807 Tau p& $+0.08\ (04)$ & $+0.14\ (05)$ & $-13.21\ (05)$ \\
V955 Tau s & V807 Tau s& $-0.02\ (04)$ & $+0.00\ (05)$ & $-13.97\ (05)$ \\
V999 Tau s & V955 Tau s& $+0.30\ (05)$ & $+0.16\ (05)$ & $-14.08\ (04)$ \\
V1026 Tau p& V807 Tau s& $+0.61\ (05)$ & $+0.20\ (04)$ & $-14.23\ (05)$ \\
V1026 Tau s& HBC 358 s & $+0.18\ (04)$ & $+0.16\ (04)$ & $-14.98\ (04)$ \\
XZ Tau p   & V807 Tau s& $+2.38\ (20)$ & $+1.30\ (15)$ & $-13.92\ (25)$ \\
XZ Tau s   & LkCa 7 s  & $+0.52\ (04)$ & $+0.06\ (03)$ & $-13.94\ (05)$ \\
\enddata
\tablenotetext{a} {Uncertainties in the last two digits appear in parentheses.}
\tablenotetext{b} {Dereddened stellar flux at 6100\AA , in units erg$\,$cm$^{-2}$s$^{-1}$\AA$^{-1}$}
\end{deluxetable}
\end{center}
   
\clearpage
\begin{center}
\begin{deluxetable}{l c c c c c c c c r r}
\singlespace
\small
\tablewidth{0 pt}
\tablenum{5}
\tablecaption{Derived properties of T Tauri Stars}
\tablehead{
\colhead{Name} & \colhead{Sp.} & \colhead{$A_V$} & \colhead{$^aL_\star$} &
\colhead{$^aL_{acc}$} & \colhead{$^b$M$_\star$} & \colhead{$^b$log($\tau$)} &
\colhead{$^c$R$_\star$} & \colhead{$^d$M$_{acc}$} &
\multispan2{$^e\Delta$K$-$L $^f\Delta$log L$_S$}
}
\startdata
DD Tau p   & M3.5 & 2.10 & 0.39 & 0.098 & 0.29&   6.17  & 1.9&$-$7.21& 0.91   &       \\
DD Tau s   & M3.5 & 2.90 & 0.24 & 0.045 & 0.29&   6.34  & 1.5&$-$7.65& 0.75   &   0.21\\
DF Tau p   & M2.0 & 0.60 & 0.52 & 0.283 & 0.38&   6.14  & 1.9&$-$6.87&\nodata &       \\
DF Tau s   & M2.5 & 0.75 & 0.68 & 0.206 & 0.35&   6.02  & 2.3&$-$6.89&\nodata &$-$0.15\\
FO Tau p   & M3.5 & 1.90 & 0.30 & 0.050 & 0.29&   6.27  & 1.6&$-$7.58& 0.43   &       \\
FO Tau s   & M3.5 & 1.90 & 0.30 & 0.031 & 0.29&   6.27  & 1.6&$-$7.78&$-$0.08 &   0.00\\
FQ Tau p   & M3.0 & 1.95 & 0.20 & 0.026 & 0.31&   6.44  & 1.3&$-$7.98& 0.10   &       \\
FQ Tau s   & M3.5 & 1.80 & 0.28 &(0.005)& 0.29&   6.28  & 1.6&($-$8.57)& 0.06 &$-$0.18\\
FS Tau p   & M0   & 4.95 & 0.15 & 0.037 & 0.61&   7.23  & 0.9&$-$8.28& 1.02   &       \\
FS Tau s   & M3.5 & 5.15 & 0.17 & 0.012 & 0.28&   6.48  & 1.2&$-$8.31& 1.18   &$-$0.72\\
FV Tau p   & K5   & 5.40 & 1.50 & 0.577 & 1.20&   6.39  & 2.1&$-$7.01& 0.20   &       \\
FV Tau s$^g$&Cont & 5.40 &\nodata&0.130 &\nodata&\nodata&\nodata&$-$7.33&0.87 &       \\ 
FV Tau/c p & M2.5 & 3.25 & 0.18 &(0.006)& 0.34&   6.56  & 1.2&($-$8.70)&0.15 &0.14$^h$\\
FV Tau/c s$^i$&M3.5&7.00 & 0.06 & 0.019 & 0.24&   6.93  & 1.1&$-$8.08& 0.83   &   0.46\\
GG Tau p   & M0   & 0.30 & 0.38 & 0.122 & 0.60&   6.61  & 1.4&$-$7.56& 0.72   &       \\
GG Tau s   & M2.0 & 0.45 & 0.20 & 0.079 & 0.38&   6.57  & 1.2&$-$7.62& 0.61   &$-$0.02\\
GH Tau p   & M2.0 & 0.00 & 0.28 & 0.027 & 0.38&   6.40  & 1.4&$-$8.02& 0.54   &       \\
GH Tau s   & M2.0 & 0.50 & 0.27 & 0.027 & 0.38&   6.42  & 1.4&$-$8.02& 0.68   &   0.02\\
HBC 358 p  & M3.5 & 0.05 & 0.12 &(0.003)& 0.27&   6.63  & 1.0&($-$8.97)&\nodata &     \\
HBC 358 s  & M4.0 & 0.05 & 0.12 &(0.003)& 0.23&   6.57  & 1.1&($-$8.86)&\nodata &$-$0.08\\
HBC 360 p  & M3.5 & 0.05 & 0.12 &(0.003)& 0.27&   6.63  & 1.0&($-$8.97)&\nodata &     \\
IS Tau p   & M0   & 3.25 & 0.64 & 0.041 & 0.58&   6.28  & 1.8&$-$7.91& 0.65   &       \\
IS Tau s   & M3.5 & 3.60 & 0.21 &(0.011)& 0.28&   6.39  & 1.4&($-$8.28)& 0.10 &   0.12\\
LkCa 7 p   & M0   & 0.20 & 0.43 &(0.015)& 0.59&   6.51  & 1.5&($-$8.44)&$-$0.07&      \\
LkCa 7 s   & M3.5 & 0.35 & 0.18 &(0.008)& 0.28&   6.45  & 1.3&($-$8.45)&$-$0.13&$-$0.06\\
UY Aur p   & M0   & 0.55 & 0.32 & 0.109 & 0.60&   6.71  & 1.3&$-$7.64& 0.81   &       \\
UY Aur s   & M2.5 & 2.65 & 0.17 & 0.064 & 0.34&   6.59  & 1.1&$-$7.70& 0.97   &$-$0.09\\
UZ Tau W p & M2.0 & 0.55 & 0.44 & 0.020 & 0.38&   6.21  & 1.8&$-$8.04& 0.68   &       \\
UZ Tau W s & M3.0 & 1.75 & 0.75 & 0.013 & 0.32&   5.93  & 2.5&$-$8.01& 0.37   &$-$0.31\\
V807 Tau p & K5   & 0.50 & 2.30 &(0.100)& 1.20&   6.13  & 2.6&($-$8.68)&0.69  &       \\
V807 Tau s & M2.0 & 0.60 & 1.04 &(0.035)& 0.38&   5.85  & 2.7&($-$8.62)&0.08  &$-$0.27\\
V927 Tau p & M3.0 & 1.40 & 0.32 &(0.011)& 0.32&   6.26  & 1.6&($-$8.28)&$-$0.16&       \\
V927 Tau s & M3.5 & 0.85 & 0.11 & 0.002 & 0.26&   6.66  & 1.0&$-$9.13&$-$0.17 &   0.44\\
V955 Tau p & K7   & 2.80 & 0.20 &(0.009)& 0.73&   7.26  & 0.9&($-$8.97)& 0.51 &       \\
V955 Tau s & M2.5 & 2.30 & 0.09 & 0.003 & 0.31&   6.92  & 0.8&$-$9.13& 0.29   &$-$0.28\\
V999 Tau p & M0.5 & 2.00 & 0.24 &(0.013)& 0.52&   6.75  & 1.1&($-$8.58)& 0.07 &       \\
V999 Tau s & M2.5 & 3.30 & 0.15 & 0.018 & 0.33&   6.65  & 1.1&$-$8.24&$-$0.26 &$-$0.08\\
V1026 Tau p& M2.0 & 2.30 & 0.09 & 0.036 & 0.35&   7.01  & 0.8&$-$8.10& 0.44   &       \\
V1026 Tau s& M3.5 & 1.90 & 0.03 & 0.002 & 0.21&   7.27  & 0.5&$-$9.34& 0.66   &   0.24\\
XZ Tau p   & M2.0 & 1.40 & 0.17 & 0.347 & 0.37&   6.66  & 1.1&$-$7.00& 1.09   &       \\
XZ Tau s   & M3.5 & 1.35 & 0.31 & 0.044 & 0.29&   6.25  & 1.7&$-$7.61& 1.61   &$-$0.45\\
&&\\
&&\\
&&\\
\enddata
\tablenotetext{a} {Dereddened stellar and excess luminosities in units of L$_\odot$; parentheses denote
upper limits.}
\tablenotetext{b} {Masses in M$_\odot$ and ages in log years measured from the pre-main-sequence
tracks of \cite{siess00}.}
\tablenotetext{c} {Stellar radii are in units of R$_\odot$.}
\tablenotetext{d} {Mass accretion rates derived from the masses, radii, and
luminosity excesses as described in the text, in units of log M$_\odot$ yr$^{-1}$.}
\tablenotetext{e} {The K$-$L excess, obtained using the IR photometry compiled by
WG01, the spectral types and reddening in this Table, the extinction law of
\cite{cardelli89}, and the intrinsic colors of \cite{kenyon95}.}
\tablenotetext{f} {Amount in log L that the secondary must move to make it coeval with the primary,
according to the tracks of \cite{siess00}.}
\tablenotetext{g} {The secondary of FV Tau is a continuum star with no photospheric lines. We adopt the
same extinction as the primary, and calculate M-dot as described in the Notes.}
\tablenotetext{h} {Amount of shift in log L required to make the primary of FV Tau/c coeval with the
primary of FV Tau.}
\tablenotetext{i} {The secondary of FV Tau/c is very faint in the blue, so the reddening is very difficult
to measure from the spectra, but must be in the range of A$_V$ $\sim$ 5 -- 8  to keep the veiling
relatively flat. The stellar luminosity is uncertain by a factor of 2. }
\end{deluxetable}
\end{center}

\clearpage
\begin{center}
\begin{deluxetable}{l c c c}
\singlespace
\small
\tablewidth{0pt}
\tabcolsep=0.3in
\tablenum{6}
\tablecaption{Dereddened Emission Line Luminosities$^a$}
\tablehead{
\colhead{Name} & \colhead{H$\alpha$} & \colhead{[O I] 6300} & \colhead{Ca II 8542} }
\startdata
DD Tau p      & $-$2.66 (0.01)& $-$4.28 (0.01) & $-$3.90 (0.04) \\
DD Tau s      & $-$2.35 (0.02)& $-$4.15 (0.01) & $-$4.59 (0.10) \\
DF Tau p$^b$  & $-$2.72 (0.06)& $-$4.35 (0.04) & $-$3.87 (0.10) \\
DF Tau s$^b$  & $-$2.69 (0.06)& $-$4.36 (0.04) & $-$4.06 (0.10) \\
FO Tau p      & $-$2.89 (0.04)& $-$5.62 (0.06) & $-$4.23 (0.08) \\
FO Tau s      & $-$3.17 (0.04)& $<-$6.3        & $<-$4.6     \\
FQ Tau p      & $-$3.17 (0.04)& $-$5.53 (0.04) & $-$4.14 (0.04) \\
FQ Tau s      & $-$3.87 (0.07)& $<-$6.7        & $<-$4.7     \\
FS Tau p      & $-$3.11 (0.03)& $-$3.90 (0.01) & $-$4.19 (0.04) \\
FS Tau s      & $-$3.80 (0.05)& $-$4.26 (0.02) & $<-$4.6     \\
FV Tau p      & $-$2.87 (0.04)& $<-$3.6        & $-$3.71 (0.10) \\
FV Tau s$^c$  & $-$3.23 (0.04)& $-$4.84 (0.08) & $-$3.80 (0.15) \\
FV Tau/c p    & $-$3.74 (0.03)& $<-$6.3        & $<-$4.8     \\
FV Tau/c s    & $-$2.27 (0.05)& $-$3.63 (0.02) & $<-$4.2     \\
GG Tau p$^b$  & $-$2.67 (0.01)& $<-$5.1        & $-$3.71 (0.06) \\
GG Tau s      & $-$3.41 (0.08)& $-$5.14 (0.06) & $<-$4.5     \\
GH Tau p      & $-$3.78 (0.05)& $<-$5.9        & $<-$4.7     \\
GH Tau s      & $-$3.80 (0.05)& $-$5.08 (0.06) & $<-$4.7     \\
HBC 358 p     & $-$4.91 (0.04)& $<-$6.9        & $<-$4.7     \\
HBC 358 s     & $-$4.81 (0.05)& $<-$6.9        & $<-$5.0     \\
HBC 360 p     & $-$4.66 (0.03)& $<-$6.9        & $<-$4.7     \\
IS Tau p      & $-$3.53 (0.02)& $-$4.76 (0.10) & $<-$4.6     \\
IS Tau s      & $-$3.89 (0.04)& $-$6.02 (0.15) & $<-$4.8     \\
LkCa 7 p      & $-$4.30 (0.03)& $<-$5.6        & $<-$4.6     \\
LkCa 7 s      & $-$4.80 (0.15)& $<-$6.8        & $<-$4.9     \\
UY Aur p      & $-$2.81 (0.04)& $-$4.42 (0.04) & $-$4.51 (0.08) \\
UY Aur s      & $-$3.10 (0.03)& $-$4.54 (0.02) & $-$3.63 (0.03) \\
UZ Tau W p    & $-$3.07 (0.06)& $<-$6.1        & $<-$4.6     \\
UZ Tau W s    & $-$2.72 (0.04)& $-$4.80 (0.04) & $<-$4.1     \\
V807 Tau p$^b$& $-$3.21 (0.04)& $<-$5.0        & $<-$4.2     \\
V807 Tau s    & $-$4.07 (0.06)& $-$4.97 (0.02) & $<-$4.8     \\
V927 Tau p    & $-$4.26 (0.06)& $<-$5.7        & $<-$4.5     \\
V927 Tau s    & $-$5.13 (0.22)& $-$5.71 (0.02) & $<-$4.5     \\
V955 Tau p    & $-$3.33 (0.02)& $<-$6.3        & $<-$4.7     \\
V955 Tau s    & $-$4.27 (0.05)& $<-$6.0        & $<-$5.0     \\
V999 Tau p    & $-$4.63 (0.07)& $<-$5.8        & $<-$4.7     \\
V999 Tau s    & $-$4.41 (0.04)& $-$4.90 (0.04) & $<-$4.4     \\
V1026 Tau p   & $-$3.36 (0.04)& $<-$6.5        & $<-$4.9     \\
V1026 Tau s   & $-$3.81 (0.02)& $-$5.52 (0.04) & $<-$4.7     \\
XZ Tau p      & $-$2.62 (0.02)& $-$4.18 (0.01) & $-$2.81 (0.02) \\
XZ Tau s      & $-$3.17 (0.04)& $-$4.99 (0.01) & $-$4.03 (0.30) \\
&&\\
&&\\
&&\\
\enddata
\tablenotetext{a} {Emission Line Luminosities in log(L/L$_\odot$), using the reddening in Table~8
and taking the distance to be 140~pc.
Line fluxes were measured from the excess spectrum with the photosphere subtracted for
Ca~II.  For [O~I] and H$\alpha$, line fluxes were measured directly from the observed spectrum.
Uncertainties tabulated reflect the error
in the equivalent width measurements, but not in the fitting of the photosphere or the
reddening.}
\tablenotetext{b} {A few pixels at the center of the H$\alpha$ line are saturated in the 2D data,
so the line luminosities listed are lower limits, but should be accurate to about 20\% based on the observed
PSFs.}
\tablenotetext{c} {Reddening assumed to be the same as that of the primary.}
\end{deluxetable}
\end{center}

\clearpage
\begin{center}
\begin{deluxetable}{l c c c c c}
\singlespace
\small
\tablewidth{6.5in}
\tablenum{7}
\tabcolsep=0.16in
\tablecaption{Accretion Signatures$^a$}
\tablehead{
\colhead{Name} & \colhead{EW(H$\alpha$)$^b$} &
\colhead{Veiling$^c$} & \colhead{IR Excess$^d$} & \colhead{[O I]$^e$} &
\colhead{Class$^f$}
}
\startdata
DD Tau p   &   X   &   X   &   X   &   X   & C \\
DD Tau s   &   X   &   X   &   X   &   X   & C \\
DF Tau p   &   X   &   X   &\nodata&   X   & C \\ 
DF Tau s   &   X   &   X   &\nodata&   X   & C \\
FO Tau p   &   X   &   X   &   X   &   X   & C \\
FO Tau s   &   X   &   X   &  $-$  &  $-$  & C \\ 
FQ Tau p   &   X   &   X   &  $-$  &   X   & C \\ 
FQ Tau s   &  $-$  &  $-$  &  $-$  &  $-$  & W \\ 
FS Tau p   &   X   &   X   &   X   &   X   & C \\
FS Tau s   &   X   &   X   &   X   &   X   & C \\
FV Tau p   &   X   &   X   &   X   &   X   & C \\ 
FV Tau s   &   X   &   X   &   X   &   X   & C \\
FV Tau/c p &  $-$  &  $-$  &  $-$  &  $-$  & W \\ 
FV Tau/c s &   X   &   X   &   X   &   X   & C \\
GG Tau p   &   X   &   X   &   X   &  $-$  & C \\
GG Tau s   &   X   &   X   &   X   &   X   & C \\
GH Tau p   &   X   &   X   &   X   &  $-$  & C \\
GH Tau s   &   X   &   X   &   X   &   X   & C \\
HBC 358 p  &  $-$  &  $-$  &\nodata&  $-$  & W \\
HBC 358 s  &  $-$  &  $-$  &\nodata&  $-$  & W \\
HBC 360 p  &  $-$  &  $-$  &\nodata&  $-$  & W \\
IS Tau p   &   X   &   X   &   X   &   X   & C \\
IS Tau s   &  $-$  &  $-$  &  $-$  &   X   & W \\
LkCa 7 p   &  $-$  &  $-$  &  $-$  &  $-$  & W \\
LkCa 7 s   &  $-$  &  $-$  &  $-$  &  $-$  & W \\
UY Aur p   &   X   &   X   &   X   &   X   & C \\
UY Aur s   &   X   &   X   &   X   &   X   & C \\
UZ Tau W p &   X   &   X   &   X   &  $-$  & C \\
UZ Tau W s &   X   &   X   &   X   &   X   & C \\
V807 Tau p &   X   &  $-$  &   X   &  $-$  & C \\
V807 Tau s &  $-$  &  $-$  &  $-$  &   X   & W \\
V927 Tau p &  $-$  &  $-$  &  $-$  &  $-$  & W \\
V927 Tau s &  $-$  &   X   &  $-$  &  $-$  & W \\
V955 Tau p &   X   &  $-$  &   X   &  $-$  & C \\
V955 Tau s &  $-$  &  $-$  &   X   &   X   & C \\
V999 Tau p &  $-$  &  $-$  &  $-$  &  $-$  & W \\
V999 Tau s &  $-$  &   X   &  $-$  &  $-$  & W \\
V1026 Tau p &  X   &   X   &   X   &  $-$  & C \\
V1026 Tau s &  X   &   X   &   X   &   X   & C \\
XZ Tau p   &   X   &   X   &   X   &   X   & C \\
XZ Tau s   &   X   &   X   &   X   &   X   & C \\
&&\\
&&\\
&&\\
\enddata
\tablenotetext{a} {The presence and absence of accretion signatures are noted with an `X' and a `$-$', respectively.}
\tablenotetext{b} {The criteria for a cTTs based on EW(H$\alpha$) depend on spectral type (see text).}
\tablenotetext{c} {Veiling detections of $>$ 2 $\sigma$ are marked with an X.}
\tablenotetext{d} {Objects with $\Delta$(K$-$L) $>$ 0.20 are marked as having excesses, 
with $\Delta$(K$-$L) defined as in Table~5.}
\tablenotetext{e} {Detections of [O I] 6300 with $>$ 2 $\sigma$ are indicated.}
\tablenotetext{e} {`W' indicates a weak-lined T Tauri star and `C' a classical T Tauri star.}
\end{deluxetable}
\end{center}


\begin{figure}
\epsscale {0.90}
\plotone{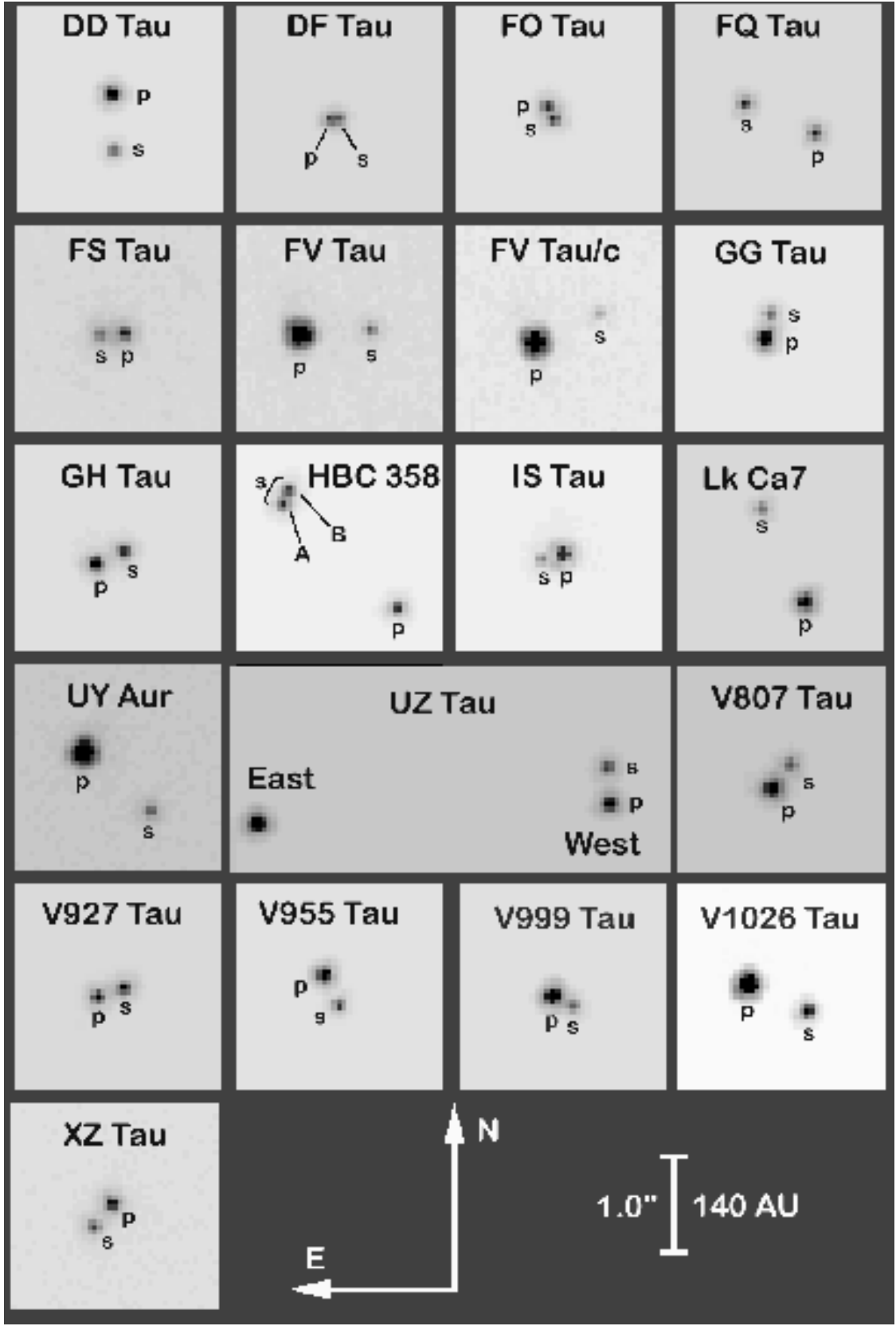}
\caption{Images of the targets in our sample obtained using the
acquisition camera on HST through a clear filter. 
\label{fig1}}
\end{figure}

\begin{figure}
\epsscale {1.20}
\plotone{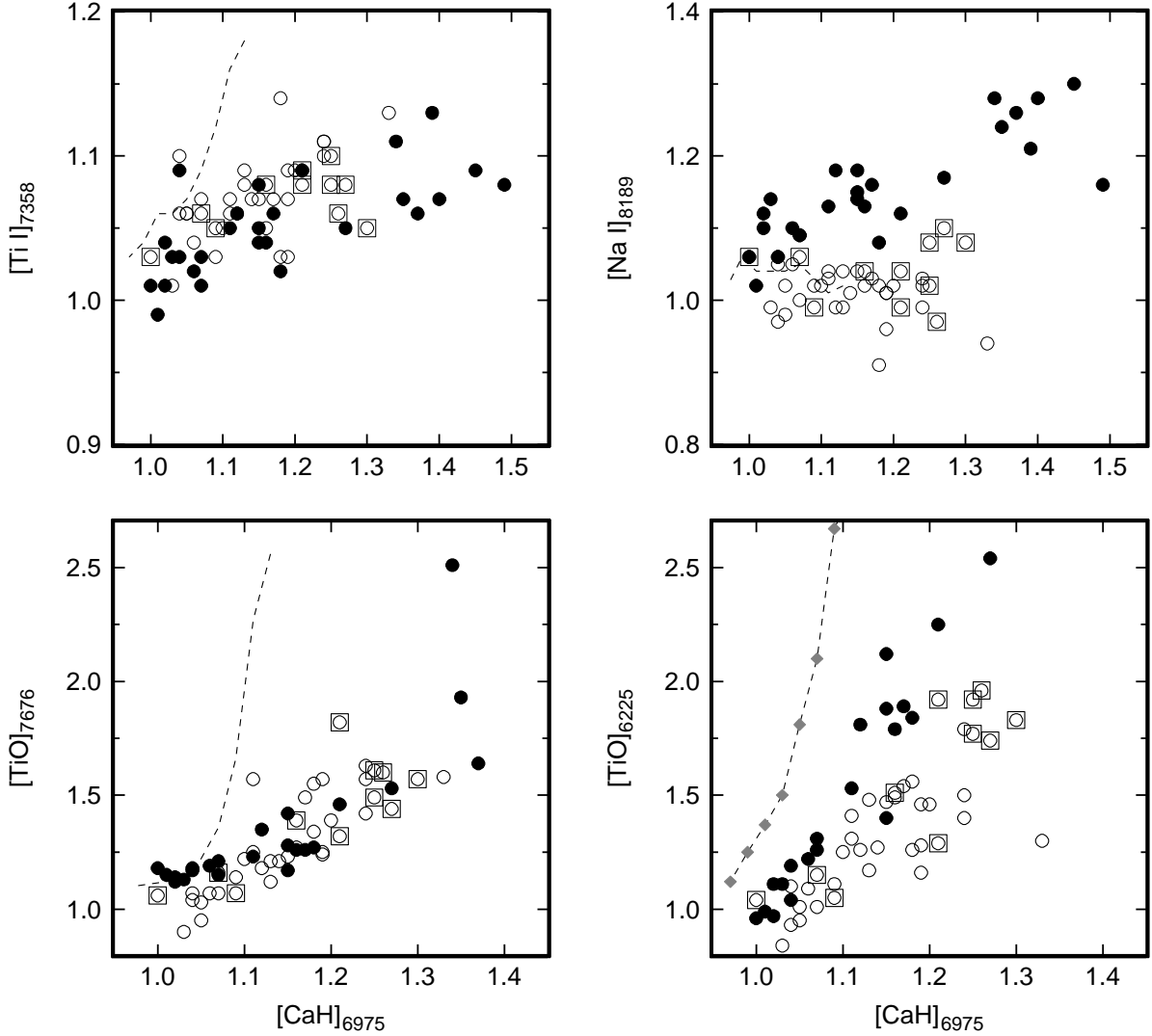}
\caption{Low resolution absorption indices.
In each panel, filled circles indicate main sequence stars,
open circles pre-main sequence stars, and
open boxes young stars without near-infrared
excess. The dashed line plots the locus of 109 M-type
giants from Kenyon \& F\'ernandez-Castro (1987).
Pre-main sequence stars follow the locus of main sequence stars
for $\rm [TiO]_{7675}$ and $\rm [Ti~I]_{7358}$, but resemble giants
in the $\rm [Na~I]_{8189}$ vs. $\rm [CaH]_{6975}$ plot.
T Tauri stars have weaker $\rm [TiO]_{6225}$ indices than do normal main
sequence stars or giants with the same $\rm [CaH]_{6975}$ index.
\label{fig2}}
\end{figure}

\begin{figure}
\def\thefigure{3}
\vbox to 5.5in{\includegraphics{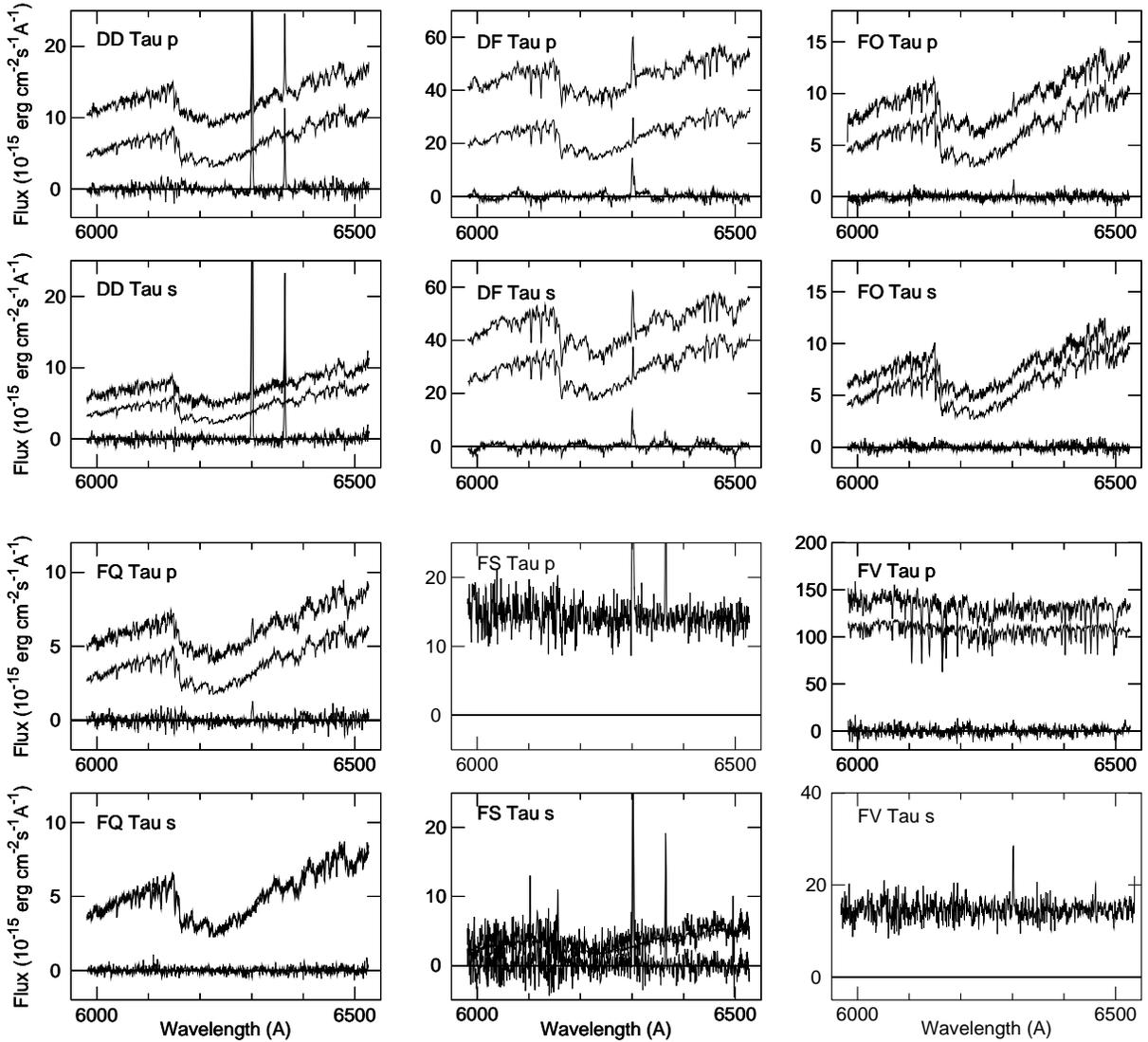}}
\caption{Medium-resolution STIS spectra of close binary T~Tauri stars.
The top curve is the dereddened spectrum of the object. The middle curve is the stellar
component, determined from fitting the depths of absorption lines in a template star
with the same spectral type as the object. Offsets between the top two
curves are caused by excess emission. Residuals between the top curve and a
model generated from the middle curve plus continuum veiling appear at the bottom
of the plots, and typically show emission lines in the object. 
The template star V807~Tau-s has [O~I]
emission, so residuals to fits that use this star sometimes show spurious absorption
at [O~I]. Reddening can differ between the primaries and their secondaries.
Stars used as photospheric templates are labeled
as such. FS Tau-p, FV Tau-s, and FV Tau/c-s are
either too faint to measure veiling in the medium-resolution data,
or have no discernible photospheric absorption lines.}
\end{figure}

\begin{figure}
\def\thefigure{3}
\vbox to 8.5in{\includegraphics{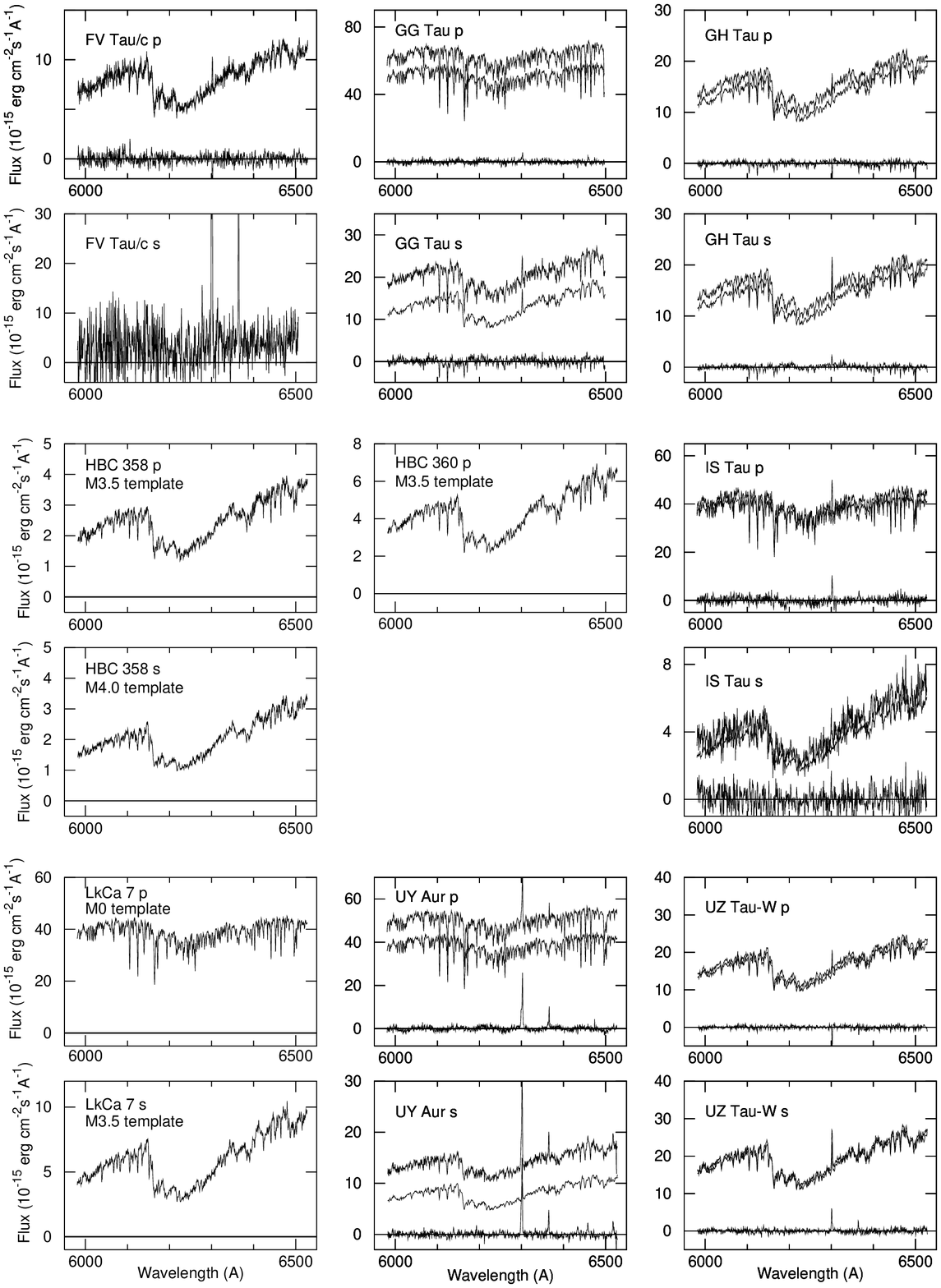}}
\caption{Cont.}
\end{figure}

\begin{figure}
\def\thefigure{3}
\vbox to 5.8in{\includegraphics{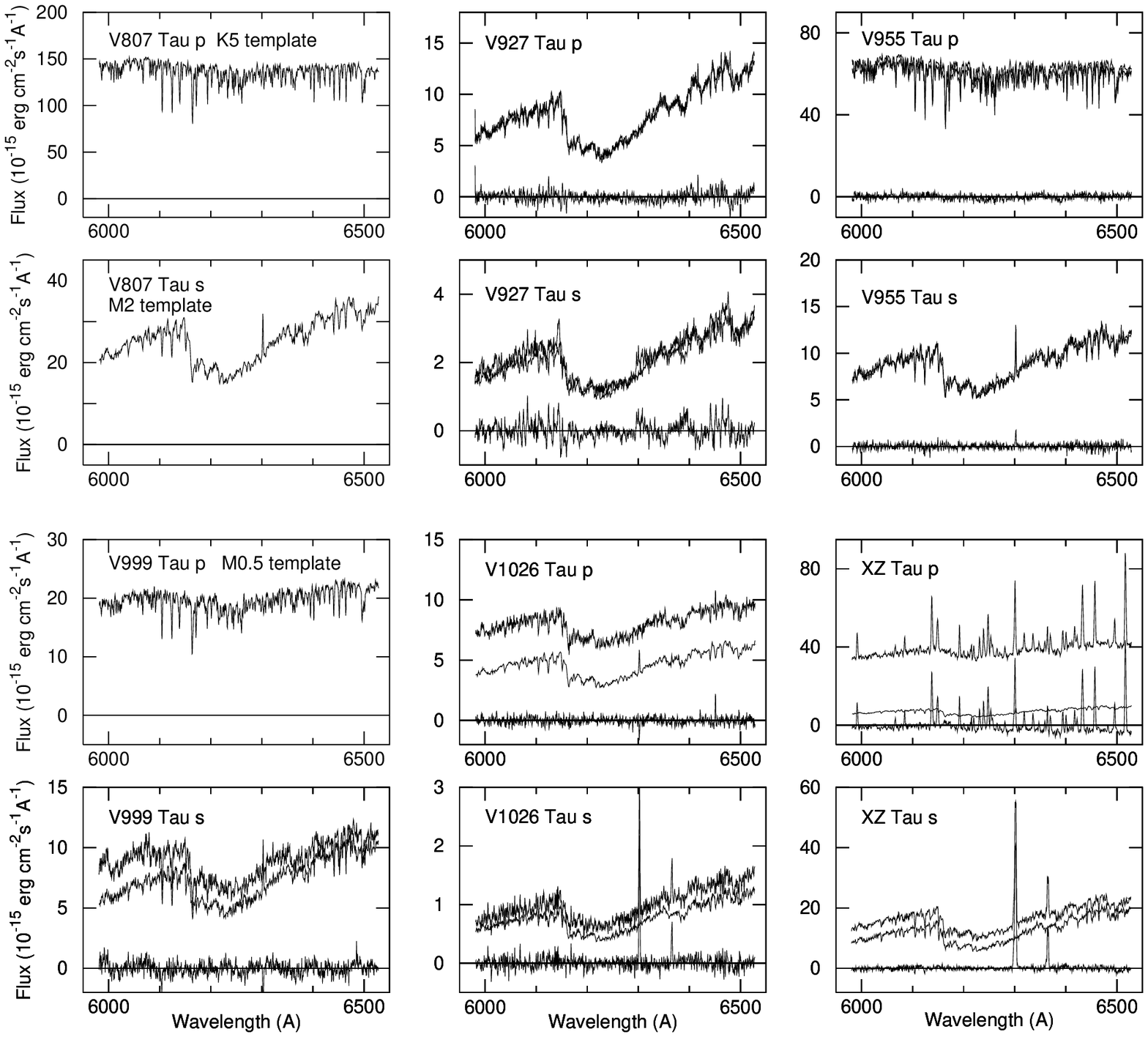}}
\caption{Cont.}
\end{figure}

\begin{figure}
\def\thefigure{4}
\vbox to 5.5in{\includegraphics{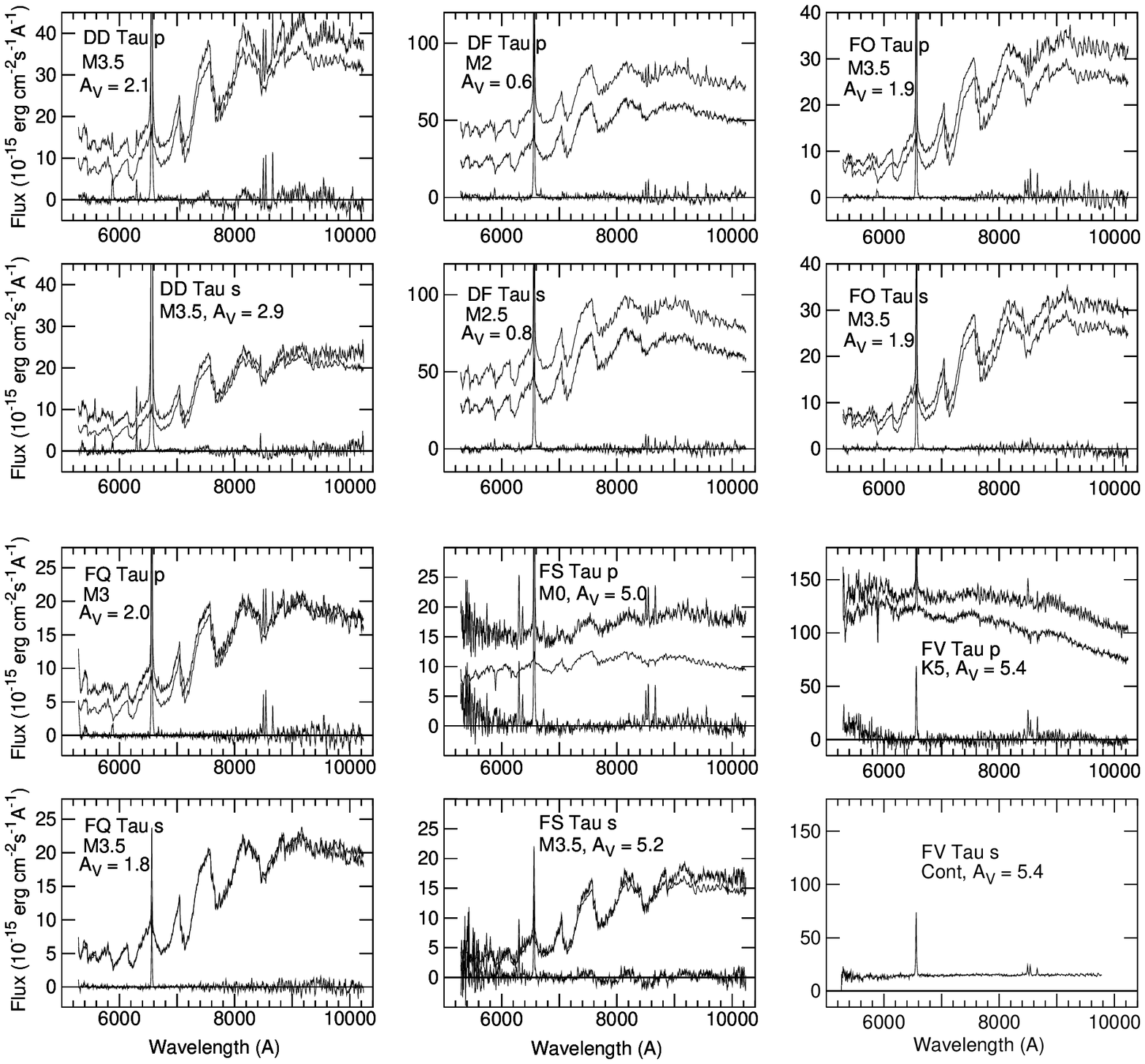}}
\caption{Same as Fig.~3 but for the dereddened low resolution STIS spectra.
The lines at 6300\AA\ and 6563\AA\ are [O~I] and H$\alpha$, respectively;
the triplet of lines near 8500\AA\ comes from Ca~II.
The y-axis scales of the secondaries are the same as their primaries in this figure.}
\end{figure}

\begin{figure}
\def\thefigure{4}
\vbox to 8.5in{\includegraphics{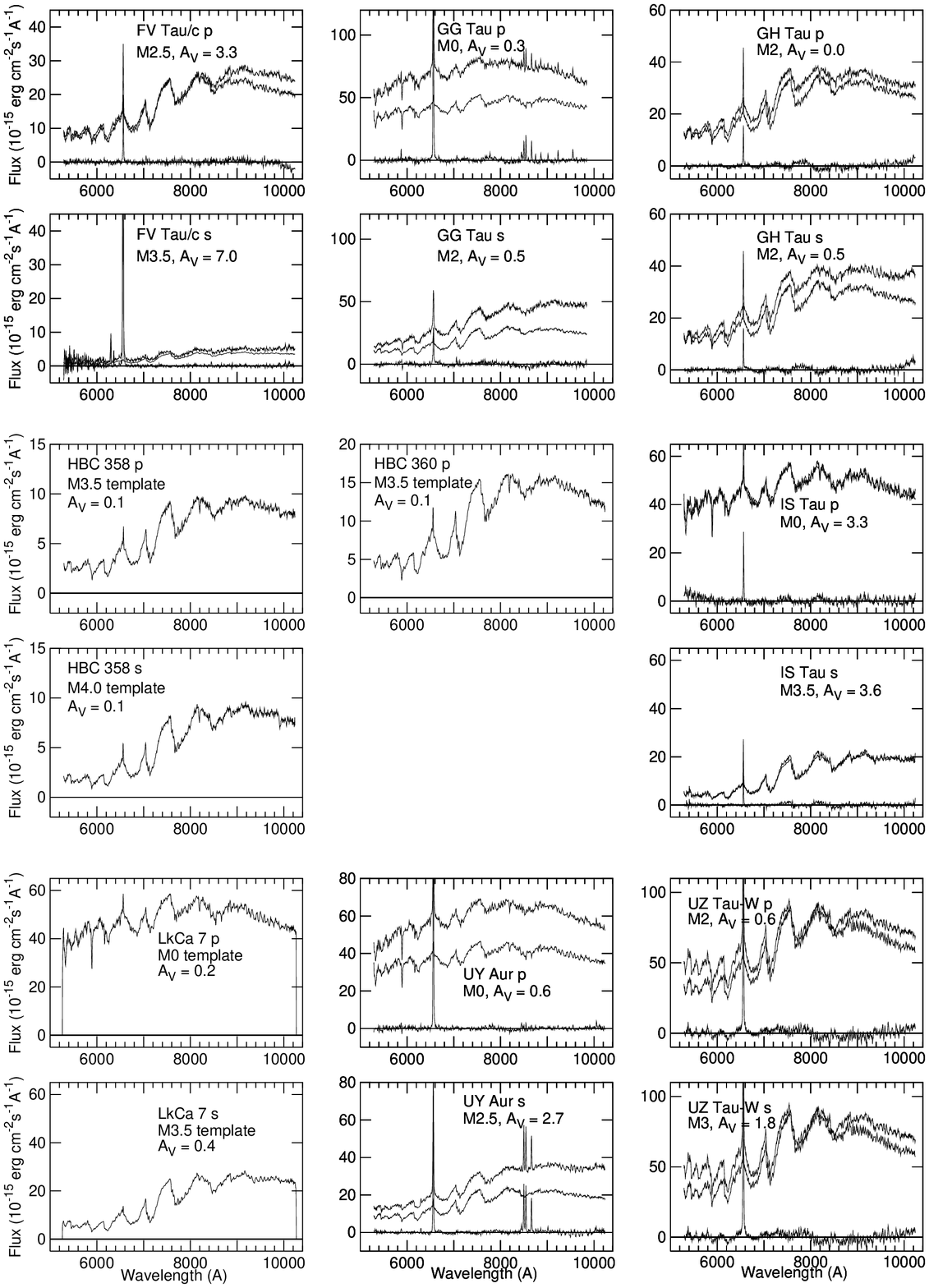}}
\caption{Cont.}
\end{figure}

\begin{figure}
\def\thefigure{4}
\vbox to 5.8in{\includegraphics{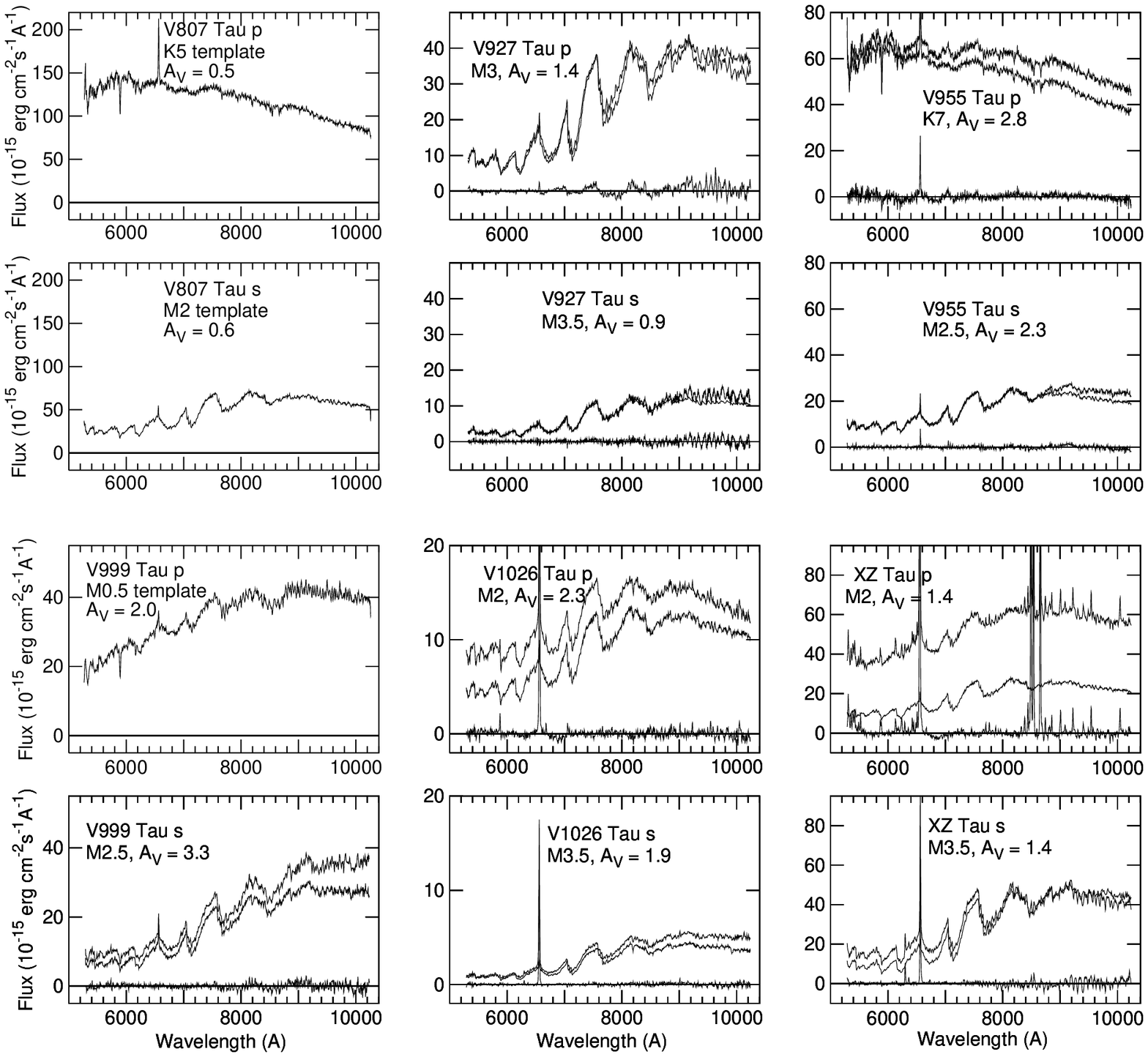}}
\caption{Cont.}
\end{figure}

\begin{figure}
\def\thefigure{5}
\vbox to 7.3in{\includegraphics{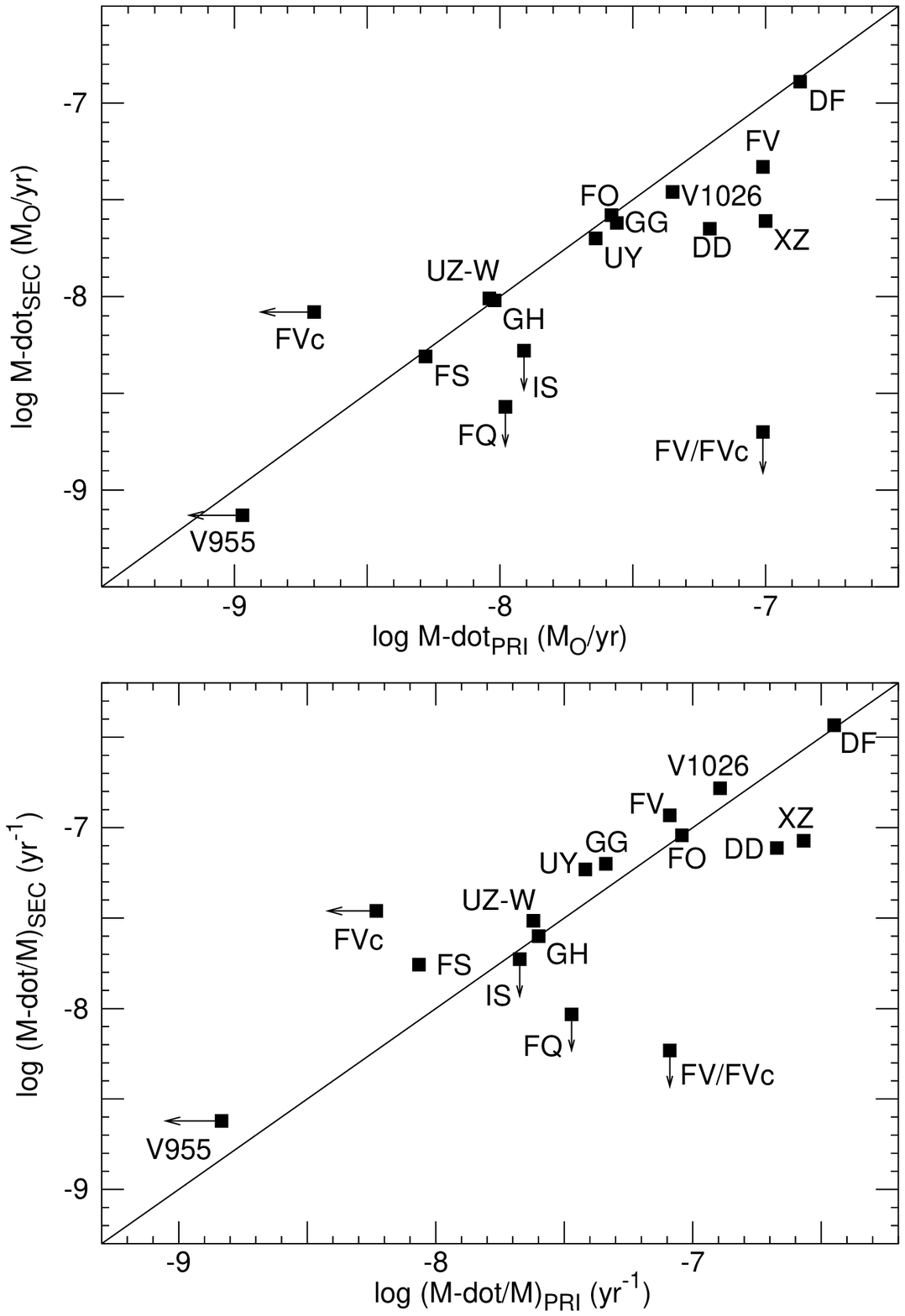}}
\caption{ Top: Mass accretion rates of secondaries and primaries.
Bottom: Accretion rates per unit mass for the secondaries and their primaries. 
The lines in both plots mark where the x- and y-values are equal.}
\end{figure}

\begin{figure}
\def\thefigure{6}
\vbox to 6.0in{\includegraphics{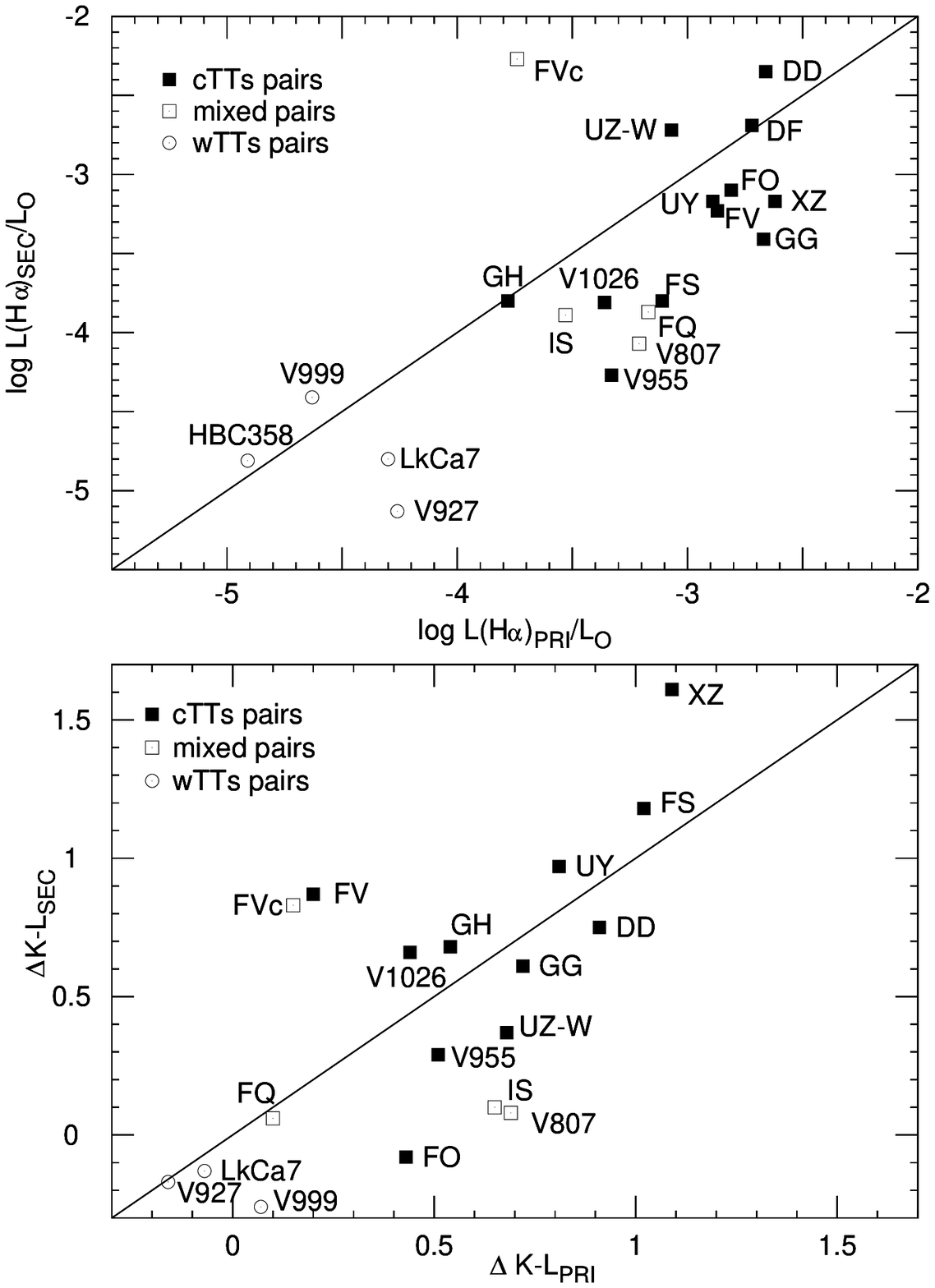}}
\caption{Top: Dereddened H$\alpha$ luminosities of the secondaries and primaries
taken from Table~6, with the classifications from Table~7. 
Points below the line have more luminous H$\alpha$ emission from the primary.
Bottom: Infrared excesses of the primary and secondary determined from
the difference between the dereddened K$-$L color and the intrinsic
K$-$L of the star inferred from its spectral type. In the case of FV~Tau,
the secondary was assumed to have an intrinsic photospheric color
of 0.16, typical of the later stars in our sample, and a reddening equal to
that of the primary. 
The line shows where the infrared excesses of the primary and
secondary are equal; points to the right of the line have more infrared color
excess from the primary.}
\end{figure}

\begin{figure}
\def\thefigure{7}
\vbox to 6.5in{\includegraphics{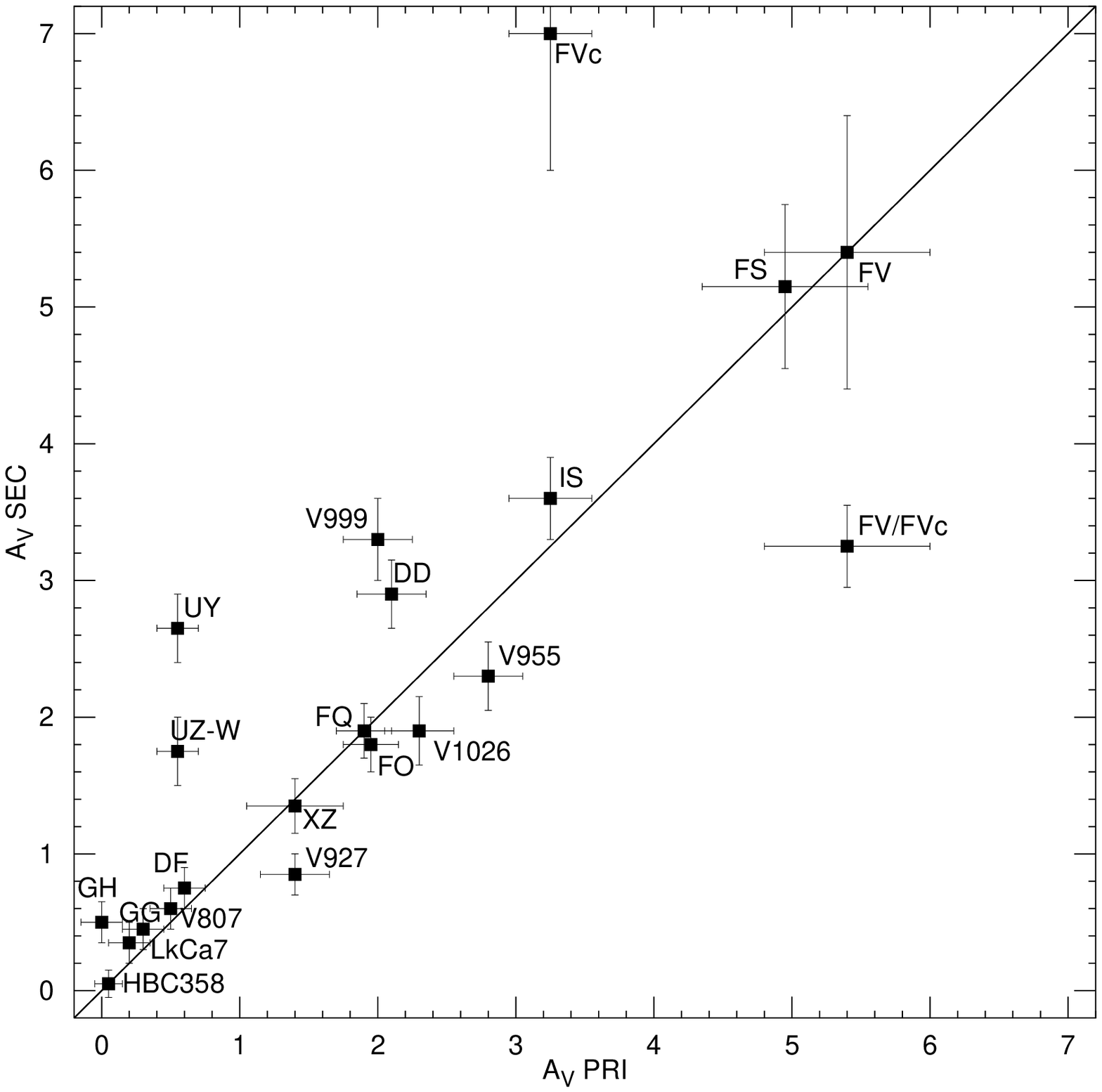}}
\caption{Extinction toward the primaries and secondaries, as determined
from spectra of each component. Extinctions are correlated, with extinctions
toward the secondary significantly exceeding those of the primary in about
30\%\ of the cases. The point labeled `FV/FVc' compares the primaries of
a wide quadruple.}
\end{figure}

\begin{figure}
\def\thefigure{8}
\vbox to 7.0in{\includegraphics{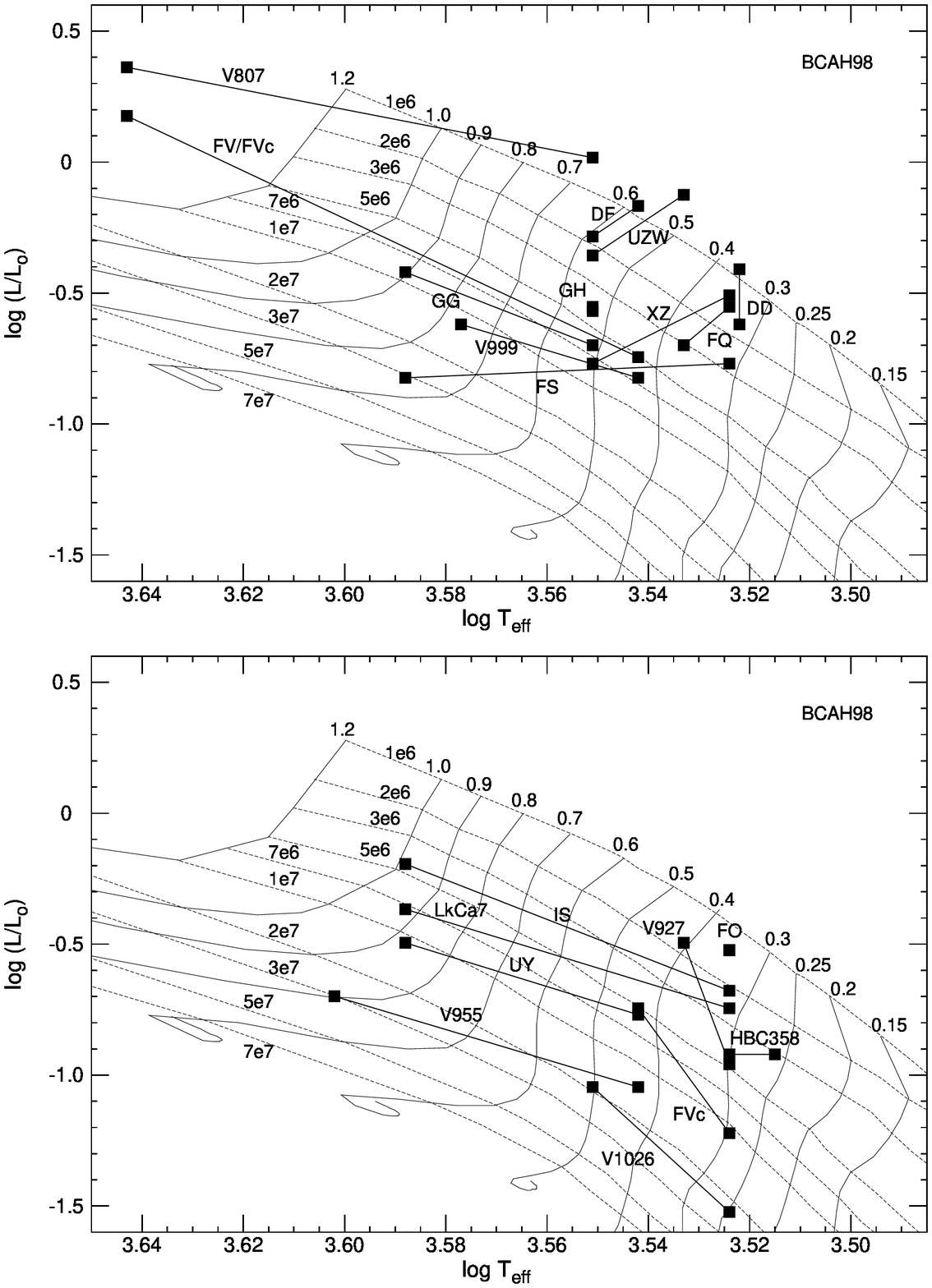}}
\caption{An HR diagram of the components of the 19 subarcsecond T~Tauri
binaries in our sample according to the models of \cite{baraffe98}. 
A solid line connects primaries and secondaries. Two panels are needed to
avoid confusion when overplotting the pairs. The less massive star
tends to be younger in several cases, and is even more luminous than
the primary in FS~Tau and XZ~Tau.}
\end{figure}

\begin{figure}
\def\thefigure{9}
\vbox to 7.0in{\includegraphics{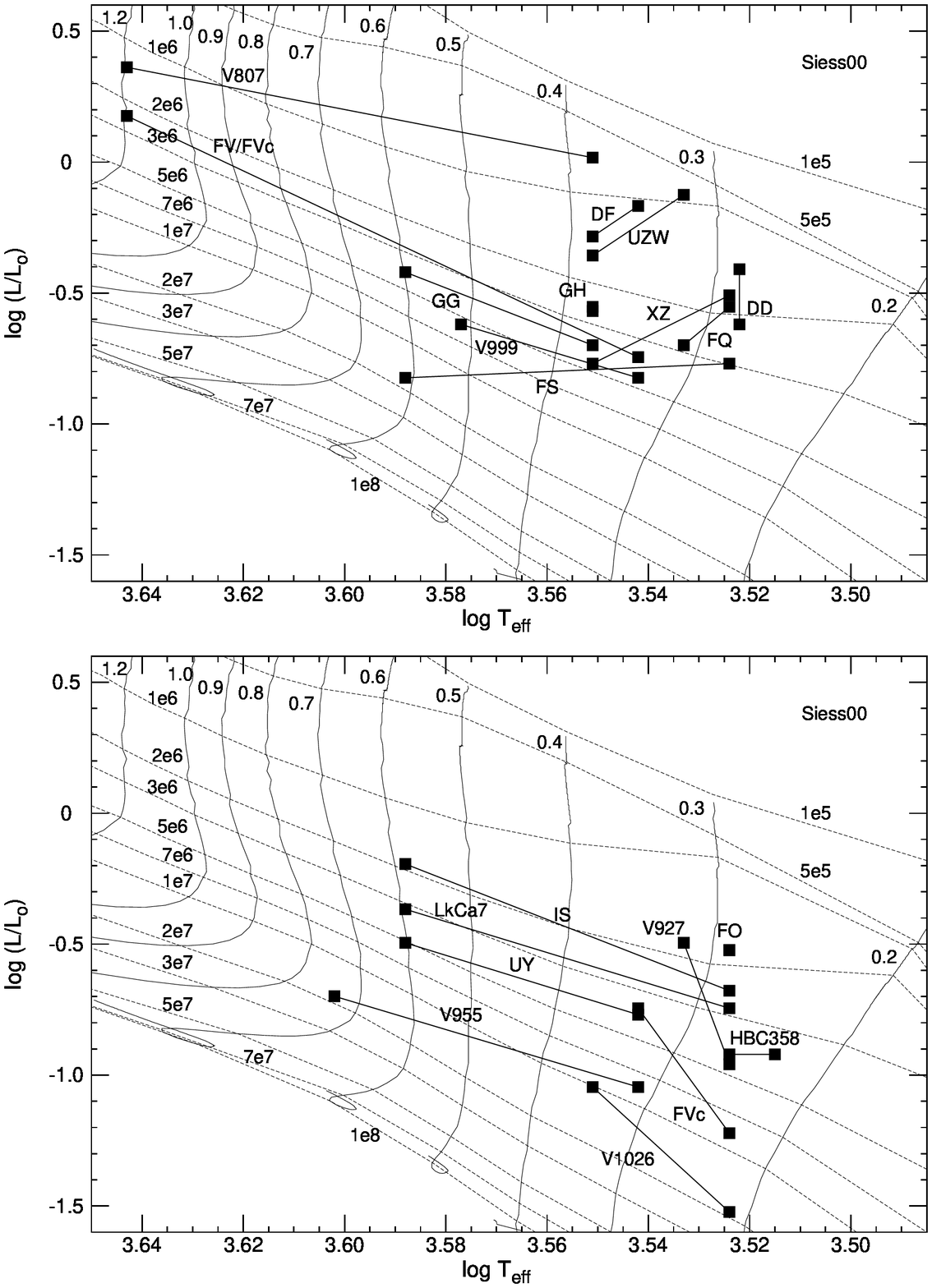}}
\caption{Same as Fig.~8 except using the pre-main-sequence tracks of
\cite{siess00}. Masses and ages derived from these tracks appear in
Table~5. We use masses and radii from these models to estimate
mass accretion rates.}
\end{figure}

\begin{figure}
\def\thefigure{10}
\vbox to 7.0in{\includegraphics{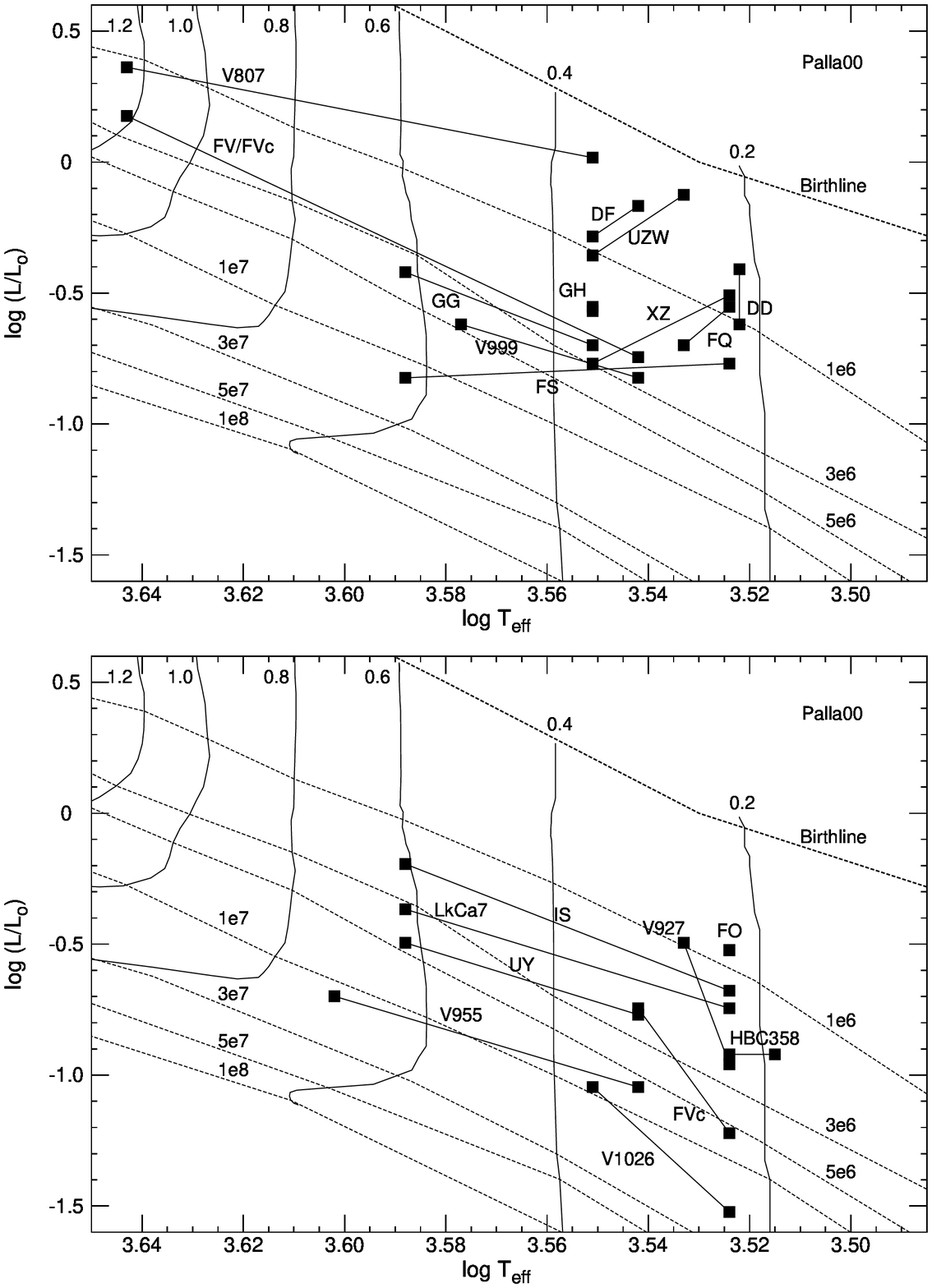}}
\caption{Same as Fig.~8 except using the pre-main-sequence tracks of
\cite{palla99}.}
\end{figure}

\begin{figure}
\def\thefigure{11}
\vbox to 7.4in{\includegraphics{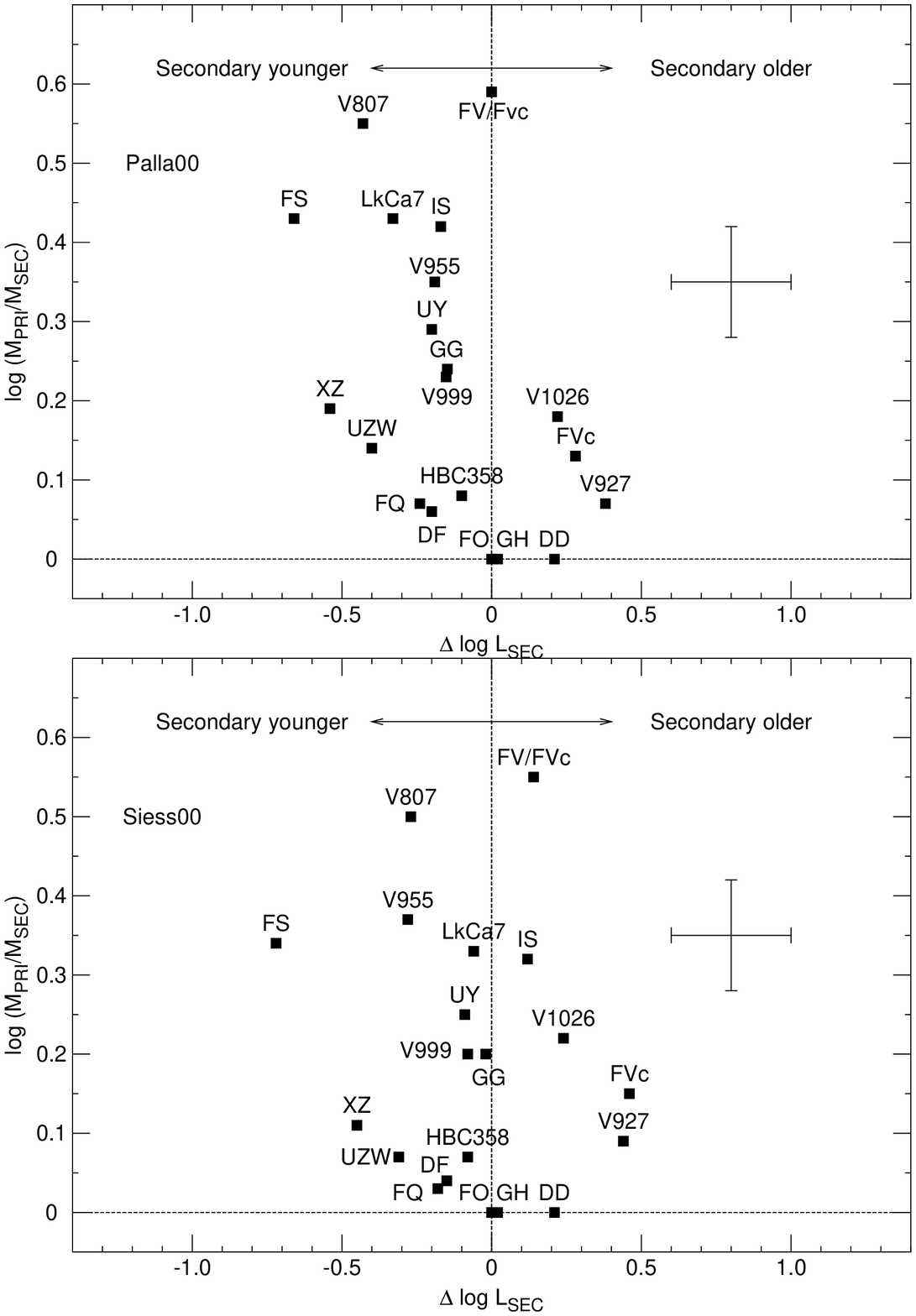}}
\caption{Binary mass ratios plotted against $\Delta\log$ L$_{SEC}$, the
amount that the secondary must move in log~L to lie on the same isochrone 
as the primary. On average, secondaries
appear slightly younger than primaries do for both sets of tracks. }
\end{figure}

\begin{figure}
\def\thefigure{12}
\vbox to 6.4in{\includegraphics{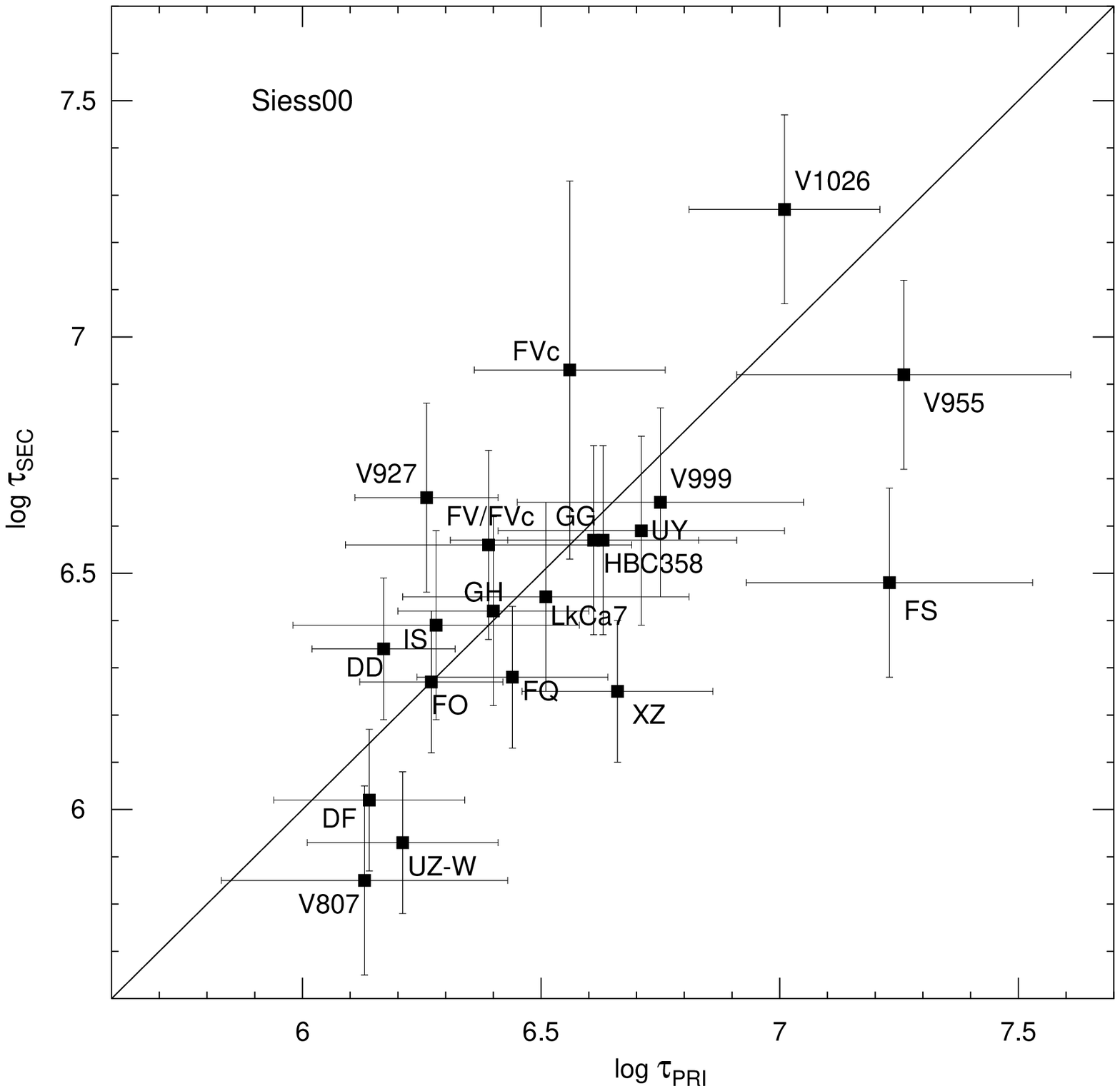}}
\caption{Ages of the primaries and secondaries in our sample according to the
tracks of \cite{siess00}. The straight line marks coeval pairs;
secondaries that lie to the right of this line are younger than their primaries
and those to the left are older than their primaries.  Ages of the primaries and
secondaries are correlated, though on average the secondaries are younger than their primaries.
Uncertainties in the ages arise from mapping the error bars in log L$_{\star}$ to
ages according to the evolutionary tracks.}
\end{figure}

\normalsize

\end{document}